\documentclass[a4paper,UKenglish,cleveref, autoref, thm-restate]{lipics-v2021}

\pdfoutput=1 
\hideLIPIcs  
\nolinenumbers 

\def\finalversion{}

\def\withappendix{}

\newcommand{\more}[1]{\ifdefined\withappendix#1\fi}

\usepackage{epsfig}
\usepackage{tikz}
\usetikzlibrary{matrix,arrows,positioning,calc,fit,backgrounds}
\usepackage{amsmath}
\usepackage{amssymb}
\usepackage{booktabs}
\usepackage[square,comma,numbers,sort&compress]{natbib}
\usepackage{listings}
\usepackage{color}
\usepackage{proof}
\usepackage{mathtools}
\usepackage[shortlabels, inline]{enumitem}
\usepackage{url}
\usepackage{inconsolata} 
\usepackage{hyperref}
\usepackage{cleveref}

\usepackage{amsfonts}
\usepackage{amssymb}
\usepackage{pftools}

\usepackage{mathpartir}

\usepackage{bm}

\usepackage{iris}

\usepackage[T1]{fontenc} 

\usepackage{tikz}
\usetikzlibrary{matrix,arrows,positioning,calc,fit,backgrounds, decorations.pathreplacing, calligraphy}

\usepackage[shortlabels, inline]{enumitem}
\makeatletter
\newcommand{\myitem}[1][]{
  \protected@edef\@currentlabel{#1}%
\item[#1]
}
\makeatother

\usepackage{mathtools}

\usepackage{etoolbox}

\usepackage{array}
\newcolumntype{L}{>$l<$}

\usepackage{stmaryrd}

\usepackage{dashbox}

\usepackage{scalerel}

\usepackage{mathpartir}

\usepackage{xspace}
\usepackage{multirow}

\usepackage{breakcites}

\usepackage{verbatim}

\usepackage{nameref}

\newcommand{\m}[1]{\mathsf{#1}}

\newcommand{\defFunc}[2]{\newcommand{#1}{\mathsf{#2}}}

\renewcommand{\le}{\leqslant}

\newcommand{\Nat}{\mathbb{N}}

\newcommand{\impl}{\Rightarrow}

\newcommand{\defeq}{\coloneqq}

\newcommand{\paren} [1] {\ensuremath{ \left( {#1} \right) }}

\newcommand{\abs}[1]{\ensuremath{\lvert #1 \rvert}}
\newcommand{\setcomp}[2]{\ensuremath{\left\{#1\;\middle|\;#2\right\}}}

\newcommand{\refFig}[1]{Figure~\ref{#1}}

\newcommand{\refRule}[1]{\ruleref{#1}}

\newcommand{\tool}[1]{\textsf{#1}}
\newcommand{\code}[1]{\textnormal{\texttt{#1}}}

\newcommand{\invariant}[2]{\boxed{#2}^{#1}}

\newcommand{\sinv}[1]{\invariant{}{#1}}

\newcommand{\ghostState}[2]{\dbox{\ensuremath{#2}}^{\,\gname_{#1}}}

\newcommand{\atomicUpdate}{\mathsf{AU}}

\makeatletter%
\@ifundefined{dplus}{%
\newcommand\dplus{\mathbin{+\kern-1.0ex+}}
}{}
\makeatother%

\newcommand{\raOp}{\cdot}

\newcommand{\authFrag}{\circ}
\newcommand{\authAuth}{\bullet}

\newcommand{\result}{\mathit{res}}

\newcommand{\mainInv}{\mathsf{Inv}}
\newcommand{\nodeInv}{\mathsf{Node}}

\defFunc{\inset}{inset}
\defFunc{\linkset}{lnks}
\defFunc{\outset}{outset}
\defFunc{\edgeset}{es}
\defFunc{\reachset}{rs}
\defFunc{\inreach}{inr}
\defFunc{\contentsFn}{cnts}
\defFunc{\keyset}{keyset}
\defFunc{\keyspace}{K}

\newcommand{\contents}{C}

\newcommand{\cssOp}{\mathtt{op}}
\newcommand{\dSpec}[1]{\Psi_{\code{#1}}}

\defFunc{\inflowFn}{inf}
\defFunc{\outflowFn}{outf}

\defFunc{\interfaceFn}{int}
\defFunc{\footprintFn}{ffp}

\defFunc{\init}{Init}
\defFunc{\capacity}{cap}
\defFunc{\capacityAux}{capAux}
\defFunc{\flowFn}{flow}
\defFunc{\flowmapFn}{flm}
\defFunc{\userEdgeFn}{edges}

\defFunc{\goodCondition}{\nu}

\defFunc{\pathCount}{pc}

\newcommand{\KS}{\mathbb{K}}

\newcommand{\inflow}{\mathit{in}}

\defFunc{\lock}{lock}
\defFunc{\pathset}{path}
\defFunc{\inflows}{In}

\defFunc{\composition}{comp}
\defFunc{\projection}{proj}

\defFunc{\abstractionFn}{edge}
\defFunc{\hrepSpatial}{spatialRep}

\newcommand{\mkblue}[1]{\textcolor{blue}{#1}}

\newcommand{\annot}[1]{\mkblue{\left\{\begin{aligned}#1\end{aligned}\right\}}}

\newcommand{\annotAtom}[1]{\mkblue{\left\langle\,\begin{aligned}#1\end{aligned}\,\right\rangle}}

\newcommand{\hoareTriple}[3]{\annot{#1} \; #2 \; \annot{#3}}

\newcommand{\atomicTriple}[3]{\annotAtom{#1} \; #2 \; \annotAtom{#3}}

\DeclareMathOperator*{\Sep}{\scalerel*{\ast}{\sum}}
\newcommand{\magicwand}{\mathrel{-\mkern-6mu*}}

\newcommand{\ite}[3]{\paren{#1 \;?\; #2 : #3}}

\newcommand{\resolve}[2]{\code{Resolve}\; {#1}\; \code{to}\; {#2}}
\newcommand{\newproph}{\code{NewProph}}
\newcommand{\proph}{\mathsf{Proph}}
\newcommand{\reg}{\mathsf{Reg}}
\newcommand{\helpingstate}{\mathsf{State}}

\newcommand{\linpending}{\mathsf{Pending}}
\newcommand{\lindone}{\mathsf{Done}}
\newcommand{\token}{\mathsf{Tok}}

\newcommand{\tid}{\mathit{tid}}

\newcommand{\paststate}{\mathsf{Past}}
\newcommand{\newthreadid}{\code{NewThreadID}}
\newcommand{\cssstate}{\mathsf{CSS}}
\newcommand{\csslstate}{\overline{\mathsf{CSS}}}

\newcommand{\cssabs}[1]{\mathsf{|}#1\mathsf{|} }
\newcommand{\pastlin}{\mathsf{PastLin}}

\newcommand{\historyinv}{\mathsf{Hist}}
\newcommand{\tplinv}{\mathsf{Inv}_{\mathit{tpl}}}
\newcommand{\helpinginv}{\mathsf{Inv}_{\mathit{help}}}

\newcommand{\thread}{\mathsf{Thread}}

\newcommand{\true}{\mathit{true}}
\newcommand{\false}{\mathit{false}}

\newcommand{\anode}{n}
\newcommand{\contentsof}[1]{\contents(#1)}
\newcommand{\keysetof}[1]{\keyset(#1)}

\newcommand{\insetof}[1]{\inset(#1)}
\newcommand{\outsetof}[1]{\outset(#1)}
\newcommand{\edgesetof}[1]{\edgeset(#1)}

\newcommand{\keyof}[1]{\m{key}(#1)}
\newcommand{\htof}[1]{\m{height}(#1)}
\newcommand{\nextof}[1]{\m{next}(#1)}
\newcommand{\markof}[1]{\m{mark}(#1)}

\DeclareFontFamily{U}{MnSymbolC}{}
\DeclareSymbolFont{MnSyC}{U}{MnSymbolC}{m}{n}
\DeclareMathSymbol{\Diamonddot}{\mathbin}{MnSyC}{"7E}
\DeclareFontShape{U}{MnSymbolC}{m}{n}{
    <-6>  MnSymbolC5
   <6-7>  MnSymbolC6
   <7-8>  MnSymbolC7
   <8-9>  MnSymbolC8
   <9-10> MnSymbolC9
  <10-12> MnSymbolC10
  <12->   MnSymbolC12}{}

\newcommand{\past}{\Diamonddot}

\lstdefinelanguage{SPL}{
  morekeywords={acc, method, struct,if,else,returns,procedure,requires,ensures,:=,var,
    new,old,free,implicit,modifies,call,locals,assume,assert,choose,havoc,ghost,
    predicate,function,invariant,while, return,atomic, split, type, field, result,
    mark, unmark, define, datatype, domain, axiom},
  deletekeywords={union,int},
  numbers=left,
  xleftmargin=2em,
  escapeinside={@}{@},
  numberstyle=\tiny,
  basicstyle=\footnotesize\ttfamily,
  columns=flexible,
  morecomment=[s][\color{green!60!black}]{/*}{*/},
  morecomment=[l][\color{green!60!black}]{//},
  moredelim=**[is][\color{purple}]{|<}{>|},
  mathescape=true,
}

\tikzset{%
  array/.style={matrix of nodes,nodes={draw, minimum size=5mm, anchor=center},column sep=-\pgflinewidth, row sep=-\pgflinewidth, nodes in empty cells,anchor=center},
  ptr/.style={*->, shorten <=-(1.8pt+1.4\pgflinewidth)},
  edge/.style={->, thick},
  dedge/.style={<->, dashed},
  fedge/.style={->, dashed},
  unode/.style={circle, draw=black, thick, minimum size=8mm, inner sep=0},
  mnode/.style={circle, draw=black, thick, fill=gray!20, minimum size=8mm, inner sep=0, font=\scriptsize},
  stackVar/.style={circle, fill=none, inner sep=0pt, minimum size=8mm, font=\normalsize, outer sep=-4pt},
  gnode/.style={circle, draw=black, thick, minimum size=8mm},
  pnode/.style={circle, draw=black, thick, minimum size=8mm},
  rnode/.style={draw=black, thick, minimum size=8mm},
  lbl/.style={circle, fill=none, inner sep=0pt, minimum size=8mm},
  dnode/.style={circle, draw=black, thick, dotted, minimum size=8mm},
  inflow/.style={circle, fill=none, inner sep=0pt, minimum size=5mm, font=\normalsize},
  phantomNode/.style={circle, fill=none, inner sep=0pt, minimum
    size=0pt}
}

\newcommand{\smartparagraph}[1]{\medskip\noindent{\textbf{#1.}}\ }

\usepackage[normalem]{ulem}

\usepackage[normalem]{ulem}

\newcommand{\defineAuthor}[3]{
  \expandafter\newcommand\csname #1\endcsname[1]{%
    \ifdefined\finalversion{##1}%
    \else{\ifdefined\monochrome{\color{green!35!black}{##1}}%
      \else{\color{#3}##1}\fi}%
    \fi}
  \expandafter\newcommand\csname #1out\endcsname[1]{%
    \ifdefined\finalversion{}%
    \else{\ifdefined\monochrome{}%
      \else{\color{#3}{\sout{##1}}}%
      \fi}%
    \fi}
  \expandafter\newcommand\csname #1footnote\endcsname[1]{%
    \ifdefined\finalversion{}%
    \else{\ifdefined\monochrome{}%
      \else{\csname #1\endcsname{\footnote{\csname #1\endcsname{#2: ##1}}}}%
      \fi}%
    \fi}
}
\defineAuthor{tw}{THOMAS}{blue!80!black}
\defineAuthor{ds}{DENNIS}{purple}
\defineAuthor{np}{NISARG}{brown!80!black}

\title{Verifying Lock-free Search Structure Templates} 

\author{Nisarg Patel}{New York University}{}{}{}
\author{Dennis Shasha}{New York University}{}{}{}
\author{Thomas Wies}{New York University}{}{}{}

\authorrunning{N. Patel, D. Shasha, and T. Wies} 

\Copyright{Nisarg Patel, Dennis Shasha, and Thomas Wies} 

\ccsdesc[500]{Theory of computation~Logic and verification}
\ccsdesc[500]{Theory of computation~Separation logic}
\ccsdesc[500]{Theory of computation~Shared memory algorithms}

\keywords{skiplists, lock-free, separation logic, linearizability, future-dependent linearization points, hindsight reasoning} 

\category{} 

\relatedversion{} 

\EventEditors{John Q. Open and Joan R. Access}
\EventNoEds{2}
\EventLongTitle{42nd Conference on Very Important Topics (CVIT 2016)}
\EventShortTitle{CVIT 2016}
\EventAcronym{CVIT}
\EventYear{2016}
\EventDate{December 24--27, 2016}
\EventLocation{Little Whinging, United Kingdom}
\EventLogo{}
\SeriesVolume{42}
\ArticleNo{23}

\begin{document}

\maketitle

\begin{abstract}
  We present and verify template algorithms for lock-free concurrent search structures that cover a broad range of existing implementations based on lists and skiplists.
  Our linearizability proofs are fully mechanized in the concurrent separation logic Iris.
  The proofs are modular and cover the broader design space of the underlying algorithms by parameterizing the verification over aspects such as the low-level representation of nodes and the style of data structure maintenance.
  As a further technical contribution, we present a mechanization of a recently proposed method for reasoning about future-dependent linearization points using hindsight arguments. The mechanization builds on Iris' support for prophecy reasoning and user-defined ghost resources.
  We demonstrate that the method can help to reduce the proof effort compared to direct prophecy-based proofs.
  \end{abstract}

\section{Introduction}

A search structure is a key-based store that implements a mutable map of keys to values (or a mutable set of keys). It provides five basic operations: (i) create an empty structure, (ii) insert a key-value pair, (iii) search for a key and return its value,  (iv) delete the entry associated with a key, and (v) update the value associated with a particular key. Because of their general usefulness, search structures are ubiquitous in data-intensive workloads.

Earlier works~\cite{DBLP:series/synthesis/2021Krishna,DBLP:journals/pacmpl/PatelKSW21,DBLP:conf/pldi/KrishnaPSW20} developed a framework to verify a wide range of lock-based implementations of concurrent search structures. Specifically, they proved that these implementations are linearizable~\cite{DBLP:journals/toplas/HerlihyW90}.

A core ingredient of the framework is the idea of template algorithms~\cite{DBLP:journals/tods/ShashaG88}.
A template algorithm dictates how threads interact but abstracts away from the concrete layout of nodes in memory. Once the template algorithm is verified, its proof can be instantiated on a variety of search structures. 


The template algorithms of~\cite{DBLP:series/synthesis/2021Krishna,DBLP:journals/pacmpl/PatelKSW21,DBLP:conf/pldi/KrishnaPSW20} use locks as a synchronization technique. Locks ensure non-interference on portions of memory to guarantee that certain needed constraints hold in spite of concurrency.

The disadvantage of locks is that if a thread holding a lock on some portion of memory $p$ stops, then no other thread can get a conflicting lock on $p$. For that reason, some practical implementations such as Java's \code{ConcurrentSkipListMap}~\cite{java-skiplist-set} use lock-free algorithms.


This paper shows how to capture multiple variants of concurrent lock-free skiplists and linked lists in the form of template algorithms. Thus, proving the correctness of such a template algorithm results in a proof that is applicable to many variants at once. Our template algorithms \np{are parametric in the skiplist height and} allow variations along the following three dimensions: 
(i) maintenance style (eager vs lazy) (ii) node implementations and (iii) the order of maintenance operations on the higher levels of the skiplists.

\np{By instantiating our template algorithm with appropriate maintenance operations and node implementations we obtain verified versions of existing (skip)list algorithms from the literature such as the Herlihy-Shavit skiplist algorithm~\cite[\S~14]{DBLP:books/daglib/0020056}, the Michael set~\cite{DBLP:conf/spaa/Michael02}, and the Harris list algorithm~\cite{DBLP:conf/wdag/Harris01}. We also obtain new concurrent skiplist algorithms that have not been considered before. These new algorithms are correct by construction thanks to our modular verification framework.}


We mechanize our development in the concurrent separation logic Iris~\cite{iris-ground-up,DBLP:conf/popl/JungSSSTBD15}. One technical contribution of our work is a formalization of {\em hindsight reasoning}~\cite{DBLP:conf/podc/OHearnRVYY10,DBLP:conf/wdag/Lev-AriCK15,DBLP:conf/wdag/FeldmanE0RS18,DBLP:journals/pacmpl/FeldmanKE0NRS20, DBLP:journals/pacmpl/MeyerWW22, DBLP:journals/pacmpl/0001W023} in Iris. Hindsight reasoning has shown its usefulness in dealing with future-dependent and external linearization points, a challenge that commonly arises in lock-free data structures.

Specifically, we build on the hindsight theory developed in~\cite{DBLP:journals/pacmpl/0001W023}, providing a mechanism in Iris where one can establish that a linearization point has passed by inferring knowledge about past states using a form of temporal interpolation.

To our knowledge, our development is the first formalization of hindsight theory in a foundational program logic. The usefulness of the developed theory extends beyond our lock-free template algorithms. In fact, we demonstrate that it can  help to reduce the proof effort compared to alternative proof techniques in Iris. To this end, we reverify the multicopy template algorithms of~\cite{DBLP:journals/pacmpl/PatelKSW21} using our formalization of hindsight as opposed to our previous tailor-made proof argument for dealing with future-dependent linearization points. The new approach reduces the proof effort by 53\%.




To summarize, our contributions are (i) template algorithms for a wide variety of lock-free search structure algorithms, (ii) mechanized proofs of linearizability based on hindsight reasoning in Iris. \np{The result is, to our knowledge, the first formal verification of fully-functional lock-free algorithms for skiplists of unbounded height.}

\section{The Skiplist Template Algorithm}
\label{sec-templates}

\newcommand{\vhead}{\mathit{hd}}
\newcommand{\vtail}{\mathit{tl}}
\newcommand{\vcurr}{\mathit{c}}
\newcommand{\vpred}{\mathit{p}}
\newcommand{\vcurrnext}{\mathit{cn}}
\newcommand{\vprednext}{\mathit{pn}}
\newcommand{\vres}{\mathit{res}}
\newcommand{\ventry}{\mathit{e}}
\newcommand{\vparr}{\mathit{ps}}
\newcommand{\vsarr}{\mathit{cs}}
\newcommand{\vperm}{\mathit{pm}}
\newcommand{\vL}{\mathsf{L}}
\newcommand{\vcurrp}{\mathit{c}'}
\newcommand{\vpredp}{\mathit{p}'}
\newcommand{\vcurrk}{\mathit{kc}}

\begin{figure}
  \centering
  \includegraphics[scale=0.5]{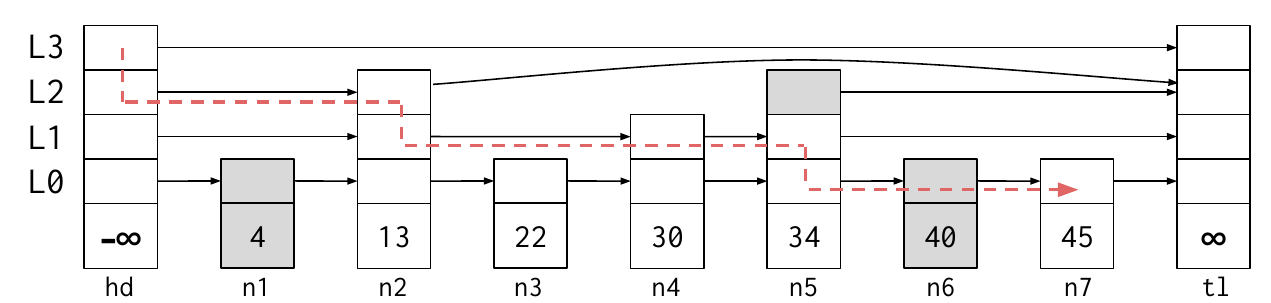}
  \caption{Skiplist with four levels. A node that is marked (logically deleted) at a level is shaded gray at that level. The red line indicates the path taken by a traversal searching for key $42$. }
  \label{fig-skiplist}
  \vspace*{-.5em}
\end{figure}

A \emph{skiplist} is a search structure over a totally ordered set of keys $\KS$.
We focus our discussion on skiplists that implement mutable sets rather than maps. The extension of the presented algorithms to mutable maps is straightforward.
The data structure is composed of sorted lists at multiple levels, with the base list determining the actual contents of the structure, while higher level lists are used to speed up the search.
An example is shown in \cref{fig-skiplist}.
A skiplist node contains a key and has a height, determining how many higher level lists this node is a part of. Each node has a next pointer for each of its levels. Two sentinel nodes signify the head ($\vhead$ with key $-\infty$) and the tail ($\vtail$ with key $\infty$) of the skiplist. 
Lock-free linked lists often use the technique of logical deletion by \emph{marking} a node before it is physically unlinked from the list. This involves storing a mark bit together with the next pointer, so as to allow reading and updating them together in a single (logically) atomic step. Lock-free skiplist implementations also use this technique. Since a skiplist node can be part of multiple lists, it has one mark bit per level. 

The traversal for a key not only goes left to right as usual, but also top to bottom. The red line in \cref{fig-skiplist} depicts a traversal searching for key $42$. The traversal begins at the highest level of the head node. At each non-base level, the traversal continues till it reaches a node with a key greater than or equal to the search key. Thereafter, the traversal drops down a level, and continues at the lower levels until it terminates on the bottom level at the first  node whose key is greater than or equal to the search key.

The traversals in a concurrent skiplist perform \emph{maintenance} in the form of physically unlinking encountered marked nodes.
In \cref{fig-skiplist}, node $n_5$ has been unlinked at level 2, thus the traversal does not visit it at that level. \np{Operations that mark and change the next pointers at the higher levels do not affect the actual contents of the structure. We therefore consider them to be part of the maintenance.}


\np{
  Many variants of lock-free skiplist algorithms have been proposed in the literature and implemented in practice. These variants differ in (i) their node implementations, (ii) the styles of maintenance operations and/or (iii) the orders in which they perform maintenance operations with regard to other operations.

For example, node implementations in low-level languages often use bit-stealing~\cite{DBLP:books/daglib/0020056} (or an equivalent of Java's \code{AtomicMarkableReference}) so that both the next pointer and mark bit can be atomically read or updated. Other implementations use more complex solutions. For instance, the skiplists in~\cite{DBLP:conf/podc/FomitchevR04} use nodes with back links to reduce traversal restarts due to marked nodes. Java's \code{ConcurrentSkipListMap}~\cite{java-skiplist-set} implements each node as a list of simpler nodes, one per level. The higher level nodes have both right pointers and down pointers, while the base nodes only have  right pointers. Java's implementation also uses \emph{marker nodes} for marking, instead of bit-stealing.

In terms of style of maintenance, the traversal in the Michael Set~\cite{DBLP:conf/spaa/Michael02} and Herlihy-Shavit lock-free skiplist~\cite[\S~14]{DBLP:books/daglib/0020056} unlinks one marked node at a time. By contrast, the traversal in the Harris List~\cite{DBLP:conf/wdag/Harris01} unlinks the entire sequence of marked nodes in one shot with a single CAS operation. The variants also differ in the order of marking of a node at higher levels. In the Herlihy-Shavit skiplist, the marking of a node goes from top level to the bottom level. This differs from skiplists in~\cite{java-skiplist-set} and~\cite{DBLP:conf/podc/FomitchevR04}, whose marking goes from bottom to top.

Despite the differences in the skiplist algorithms described above  (and others to be invented in the future), the bulk of their correctness reasoning remains the same. A goal of this paper is to show how to exploit that fact. 

\smartparagraph{Template algorithm}
Our template algorithm for skiplists abstracts away from node-level implementation details and the way in which traversals perform maintenance. As we shall see, the particular details regarding how the data is stored internal to the node does not affect the correctness of the core operations - \code{search}, \code{insert} and \code{delete}. Nor is the correctness affected by whether the traversal unlinks one marked node at a time or an entire sequence of marked nodes. We also show that the order in which maintenance operations are performed on the higher levels of the list does not matter for correctness. In summary, the template algorithm we present abstracts from: (i) node-level details; (ii) the style of unlinking marked nodes and (iii) the order of maintenance operations on higher levels.}

The template algorithm is assumed to be operating on a set of nodes $N$ that contains the two sentinel nodes head $\vhead$ and tail $\vtail$. Let the maximum allowed height of a skiplist node be $\vL$ ($> 1$). Each node $n$ is associated with (i) its key $\keyof{n} \in \KS = \Nat \cup \{-\infty, \infty\}$, (ii) its height $\htof{n} \in [1,L)$ , (iii) the next pointers $\nextof{n, i} \in N$ for each $i$ from $0$ to $\htof{n} - 1$, and (iv) its mark bits per level $\markof{n, i} \in \{\true, \false\}$ for each $i$ from $0$ to $\htof{n} - 1$. When discussing $\nextof{n, i}$ or $\markof{n, i}$, we implicitly assume that $i$ lies between $0$ and $\htof{n} - 1$.
We sometimes say a node $n$ is unmarked to mean that it is unmarked at the base level, i.e., $\markof{n, 0} = \false$. 
The structural invariant maintains the following facts: $\keyof{\vhead} = -\infty$, $\keyof{\vtail} = \infty$, $\htof{\vhead} = \htof{\vtail} = \vL$, $\nextof{\vtail, i} = \vtail$ for all $i$, $\nextof{\vhead, \vL-1} = \vtail$, $\markof{\vhead, i} = \markof{\vtail, i} = \false$ for all $i$.

The core operations of the skiplist template are expressed using \emph{helper functions} such as \lstinline+|<findNext>|+ and \lstinline+|<markNode>|+ that abstract from the details of the node implementation. We describe the behavior of these helper functions as and when we encounter them. The template is instantiated by implementing these functions. The helper functions are assumed to be \emph{logically atomic}, i.e., appear to take effect in a single step during its execution.

\Cref{fig:skiplist-template} shows the core operations of the skiplist template algorithm. (We omit the code for the data structure initialization as it is straightforward.)
All three operations begin by allocating two arrays $\vparr$ and $\vsarr$ via \code{allocArr}, each of size $\vL$ and values initialized to $\vhead$ and $\vtail$ respectively. These arrays are then populated by the \code{traverse} operation as it computes the predecessor-successor pair for operation key $k$ at each level.
Intuitively, these pairs indicate where $k$ would be inserted at each level.
The template algorithm here abstracts away from the concrete \code{traverse} implementation. We later consider two implementations of \code{traverse} that differ in the way that maintenance is performed, as discussed earlier.

As far as the core operations are concerned, they rely on \code{traverse} to satisfy the following specification. 
First, it returns a triple $(\vpred, \vcurr, \vres)$ where $\vpred$ and $\vcurr$ are nodes and $\vres$ a Boolean such that $\vpred = \vparr[0]$, $\vcurr = \vsarr[0]$ and $\vres$ is true iff $k$ is contained in $\vcurr$. Second, the node $\vcurr$ must have been unmarked at some point during the traversal; and third, for each $0 \le i < \vL$, the traversal observes that $\keyof{\vparr[i]} < k \le \keyof{\vsarr[i]}$.



\begin{figure}[t]
  \centering
  \begin{minipage}[t]{.50\textwidth}
\begin{lstlisting}[aboveskip=0pt,belowskip=0pt, frame=none]
let search $k$ =
  let $\vparr$ = allocArr $\vL$ $\vhead$ in
  let $\vsarr$ = allocArr $\vL$ $\vtail$ in
  let _, _, $\vres$ = traverse $\vparr$ $\vsarr$ $k$ in
  $\vres$

let delete $k$ =
  let $\vparr$ = allocArr $\vL$ $\vhead$ in
  let $\vsarr$ = allocArr $\vL$ $\vtail$ in
  let $\vpred$, $\vcurr$, $\vres$ = traverse $\vparr$ $\vsarr$ $k$ in@\label{line-delete-traverse}@
  if not $\vres$ then
    $\false$
  else
    maintainanceOp_del $\vcurr$; 
    match |<markNode>| 0 $\vcurr$ with @\label{line-delete-markNode}@
    | Success -> traverse $\vparr$ $\vsarr$ $k$; $\true$
    | Failure -> $\false$
\end{lstlisting}
\end{minipage}%
\begin{minipage}[t]{.50\textwidth}
\begin{lstlisting}[aboveskip=0pt,belowskip=0pt, frame=none, firstnumber=last]
let insert $k$ =
  let $\vparr$ = allocArr $\vL$ $\vhead$ in
  let $\vsarr$ = allocArr $\vL$ $\vtail$ in
  let $\vpred$, $\vcurr$, $\vres$ = traverse $\vparr$ $\vsarr$ $k$ in
  if $\vres$ then
    $\false$
  else
    let $h$ = randomNum $\vL$ in
    let $\ventry$ = |<createNode>| $k$ $h$ $\vsarr$ in
    match |<changeNext>| 0 $\vpred$ $\vcurr$ $\ventry$ with@\label{line-insert-changeNext}@
    | Success -> 
      maintainanceOp_ins $k$ $\vparr$ $\vsarr$ $\ventry$; $\true$
    | Failure -> insert $k$ 
		
\end{lstlisting}
\end{minipage}
\caption{\label{fig:skiplist-template} The template algorithm for lock-free skiplists. The template can be instantiated by providing implementations of \code{traverse} and the helper functions \code{markNode}, \code{createNode} and \code{changeNext}. The \code{markNode}$\,i\;\vcurr$ attempts to mark node $\vcurr$ at level $i$ atomically, and fails if $\vcurr$ has been marked already. \code{createNode}$\,k\,h\,\vsarr$ creates a new node $\ventry$ of height $h$ containing $k$, and whose next pointers are set to nodes in array $\vsarr$. Finally, \code{changeNext}$\,i\,\vpred\,\,\vcurr\,\,\vcurrnext$ is a CAS operation attempting to change the next pointer of $\vpred$ from $\vcurr$ to $\vcurrnext$. \code{changeNext}$\,i\,\vpred\,\,\vcurr\,\,\vcurrnext$ succeeds only if $\markof{\vpred, i} = \false$ and $\nextof{\vpred, i} = \vcurr$. Other functions used here include \code{randomNum} to generate a random number and maintenance operations associated with \code{insert} and \code{delete}. \code{maintainanceOp\_del} marks node $\vcurr$ at the higher levels, while \code{maintainanceOp\_ins} inserts a new node $\ventry$ at the higher levels.}
\end{figure}

Let us now describe the core operations, starting with the \code{search} operation. The \code{search} operation simply invokes the \code{traverse} function, whose result establishes whether $k$ was in the structure. The \code{delete} operation starts similarly by invoking \code{traverse} and checking if the key is present in the structure. If it is, then \code{delete} invokes the maintenance operation \code{maintainanceOp\_del}, which attempts to mark $\vcurr$ at the higher levels (i.e. all levels except $0$). We provide the implementation of \code{maintainanceOp\_del} in a moment. Once \code{maintainanceOp\_del} terminates, \code{delete} finally attempts to mark $\vcurr$ via \code{markNode} at the base level. If marking succeeds, it terminates by invoking \code{traverse} (which performs the task of physically unlinking marked nodes at all levels) and returning $\true$. Otherwise, a concurrent thread must have already marked $\vcurr$, in which case \code{delete} returns $\false$.

The \code{insert} operation also begins with \code{traverse}. If the traversal returns $\true$, then the key must already have been present. Hence, \code{insert} returns $\false$ in this case. Otherwise, a new node $\ventry$ is created using \code{createNode}. The node's height is determined randomly using $\code{randomNum}$, which generates a random number $h$ such that $0 < h < L$. After creating a new node, the algorithm attempts to insert it into the list by calling \code{changeNext} at the base level (line~\ref{line-insert-changeNext}). If the attempt succeeds, \code{insert} proceeds by invoking the maintenance operation \code{maintainanceOp\_ins}, which also inserts the new node into the list at all higher levels.  The \code{insert} then returns with $\true$. If the \code{changeNext} operation fails, then the entire operation is restarted.

We now describe the maintenance operations for \code{insert} and \code{delete}, shown in \Cref{fig:skiplist-maintenance}. The maintenance operations here differ from those in traditional skiplist implementations in regards to the order in which maintenance is performed at higher levels. In traditional implementations, the marking of a node goes from top to bottom, while insertion of a new node goes from bottom to top. The skiplist template presented here makes sure that the base level gets marked at the end and the insertion first happens at the base level, but it imposes no order on how it proceeds at higher levels. That is, when marking a node, a \code{delete} thread could for instance first mark odd levels, then even levels and finally the base level $0$. The maintenance operations in the skiplist template captures all such permutations. As our proof shows later, the order of maintenance at higher levels has no bearing on the correctness of the algorithm. 

\begin{figure}[h]
  \centering
  \begin{minipage}[t]{.50\textwidth}
\begin{lstlisting}[aboveskip=0pt,belowskip=0pt, frame=none]
let maintainanceOp_del_rec $i$ $h$ $\vperm$ $\vcurr$ = 
  if $i$ < $h$-1 then
    let $idx$ = $\vperm$[$i$] in
    |<markNode>| $idx$ $\vcurr$;@\label{line-delete-mnt-markNode}@
    maintainanceOp_del_rec ($i$+1) $h$ $\vperm$ $\vcurr$
  else ()

let maintainanceOp_del $\vcurr$ = 
  let $h$ = |<getHeight>| $\vcurr$ in
  let $\vperm$ = permute $h$ in
  maintainanceOp_del 0 $h$ $\vperm$ $\vcurr$
\end{lstlisting}
\end{minipage}%
\begin{minipage}[t]{.60\textwidth}
\begin{lstlisting}[aboveskip=0pt,belowskip=0pt, frame=none, firstnumber=last]
let maintainanceOp_ins_rec $i$ $h$ $\vperm$ $\vparr$ $\vsarr$ $\ventry$ = 
  if $i$ < $h$-1 then
    let $idx$ = $\vperm$[$i$] in
    let $\vpred$ = $\vparr$[$idx$] in
    let $\vcurr$ = $\vsarr$[$idx$] in
    match |<changeNext>| $idx$ $\vpred$ $\vcurr$ $\ventry$ with
    | Success -> 
      maintainanceOp_ins_rec ($i$+1) $h$ $\vperm$ $\vparr$ $\vsarr$ $\ventry$
    | Failure -> 
      traverse $\vparr$ $\vsarr$ $k$;
      maintainanceOp_ins_rec $i$ $h$ $\vperm$ $\vparr$ $\vsarr$ $\ventry$
  else ()

let maintainanceOp_ins $k$ $\vparr$ $\vsarr$ $\ventry$ = 
  let $h$ = |<getHeight>| $\ventry$ in
  let $\vperm$ = permute $h$ in
  maintainanceOp_ins 0 $h$ $\vperm$ $\vparr$ $\vsarr$ $\ventry$
\end{lstlisting}
\end{minipage}
\caption{\label{fig:skiplist-maintenance} The maintenance operations for the skiplist. The $\code{getHeight}\,\vcurr$ helper function returns $\htof{\vcurr}$. The \code{permute} function generates a permutation of $[1 \dots (h-1)]$ as an array.}
\end{figure}

The \code{maintainanceOp\_del} marks node $\vcurr$ from levels $1$ to $\htof{\vcurr}$. It begins by reading the height of $\vcurr$ as $h$, and generating a permutation of $[1 \dots (h-1)]$ stored in array $\vperm$ via the \code{permute} function. The \code{maintainanceOp\_del\_rec} then recursively marks $\vcurr$ in the order prescribed by $\vperm$. Note that the maintenance continues regardless of whether \code{markNode} succeeds or fails, because $\vcurr$ will be marked at the end regardless.

The \code{maintainanceOp\_ins} begins in the same way by reading the height, generating the permutation and invoking \code{maintainanceOp\_ins\_rec}. The \code{maintainanceOp\_ins\_rec} first collects the predecessor-successor pair at the current level from arrays $\vparr$ and $\vsarr$, respectively. Then it tries to insert the new node $\ventry$ using \code{changeNext} on predecessor node $\vpred$. If \code{changeNext} succeeds, then the recursive operation continues. Otherwise, it recomputes the predecessor-successor pairs using \code{traverse}. After the recomputation, the insertion is retried at the same level. 

We can now finally turn to the implementations of \code{traverse}. We consider two implementations that differ in their treatment of marked nodes. The \emph{eager} traversal attempts to unlink every marked node it encounters, while the \emph{lazy} traversal simply walks over the marked nodes till it reaches an unmarked node. The traversal then attempts to unlink the entire marked segment at once. The two implementations are similar in other aspects, so we  discuss only the eager traversal in detail here. \more{The details about the lazy traversal can be found in \cref{sec-lazy-traverse}.}

\newcommand{\eagerRec}{\code{eager\_rec}}
\newcommand{\eagerI}{\code{eager\_i}}

The eager traversal is shown in \cref{fig:eager-traversal}. The \code{traverse} function is implemented using mutually-recursive functions $\eagerRec$ and $\eagerI$\footnote{For ease of exposition, the implementation of the eager traversal shown in \cref{fig:eager-traversal} differs slightly from the version we have verified in Iris. The Iris version uses option return types instead of mutually-recursive functions in order to obtain a more modular proof of the eager traversal. We use the mutually recursive implementation here for clarity of exposition.}. The function $\eagerRec$ populates the arrays $\vparr$ and $\vsarr$ with the predecessor-successor pair at level $i$ computed by $\eagerI$. The $\eagerI$ performs a traversal at level $i$ by first reading the mark bit and next pointer of $\vcurr$ using \code{findNext}. If $\vcurr$ is found to be marked, then $\eagerI$ attempts to physically unlink the node using \code{changeNext}. In the case that \code{changeNext} fails (because either $\vpred$ is marked or it does not point to $\vcurr$ anymore), $\eagerI$ simply restarts the \code{traverse} function. In the case of \code{Success} of \code{changeNext}, the traversal continues. If $\vcurr$ is unmarked, then \code{traverse\_i} proceeds by comparing $k$ to $\keyof{\vcurr}$. For $\keyof{\vcurr} < k$, the traversal continues with $\vcurr$ and $\vcurrnext$. Otherwise, $\eagerI$ ends at $\vcurr$, returning $(\vpred, \vcurr, \true)$ if $\keyof{\vcurr} = k$ and $(\vpred, \vcurr, \false)$ otherwise. As mentioned before, $\eagerI$ attempts to unlink immediately whenever a marked node is encountered. 

\begin{figure}[t]
  \centering
  \begin{minipage}[t]{.55\textwidth}
\begin{lstlisting}[aboveskip=0pt,belowskip=0pt, frame=none]
let $\eagerI$ $i$ $k$ $\vpred$ $\vcurr$ =   
  match |<findNext>| $i$ $\vcurr$ with@\label{eager-tri-findNext}@
  | $\vcurrnext$, $\true$ -> 
    match |<changeNext>| $i$ $\vpred$ $\vcurr$ $\vcurrnext$ with
    | Success -> $\eagerI$ $i$ $k$ $\vpred$ $\vcurrnext$ @\label{eager-rec1}@
    | Failure -> traverse $\vparr$ $\vsarr$ $k$
  | $\vcurrnext$, $\false$ ->@\label{eager-tri-findNext-unmarked}@
    let $\vcurrk$ = |<getKey>| $\vcurr$ in
    if $kc$ < $k$ then
      $\eagerI$ $i$ $k$ $\vcurr$ $\vcurrnext$ @\label{eager-rec2}@
    else 
      let $\vres$ = ($kc$ = $k$ ? $\true$ : $\false$) in
      ($\vpred$, $\vcurr$, $\vres$)@\label{eager-tri-return}@
\end{lstlisting}
\end{minipage}%
\begin{minipage}[t]{.50\textwidth}
\begin{lstlisting}[aboveskip=0pt,belowskip=0pt, frame=none, firstnumber=last]
let $\eagerRec$ $i$ $\vparr$ $\vsarr$ $k$ =
  let $\vpred$ = $\vparr$[$i$+1] in
  let $\vcurr$, _ = |<findNext>| $i$ $\vpred$ in@\label{eager-rec-findNext}@
  let $\vpredp$, $\vcurrp$, $\vres$ = $\eagerI$ $i$ $k$ $\vpred$ $\vcurr$ in@\label{eager-rec-eageri}@
  $\vparr$[$i$] <- $\vpredp$;
  $\vsarr$[$i$] <- $\vcurrp$;
  if $i$ = 0 then
    ($\vpredp$, $\vcurrp$, $\vres$)
  else
  $\eagerRec$ ($i$-1) $\vparr$ $\vsarr$ $k$

let traverse $\vparr$ $\vsarr$ $k$ =
  $\eagerRec$ ($\vL$ - 2) $\vparr$ $\vsarr$ $k$
\end{lstlisting}
\end{minipage}
\caption{\label{fig:eager-traversal} The eager traversal for the skiplist template.
  \code{findNext}$\,i\,k\,\,\vcurr$ returns a pair $(\nextof{\vcurr, i}, \markof{\vcurr, i})$. The $\code{getKey}\,\vcurr$ helper function returns $\keyof{\vcurr}$.}
\end{figure}


\section{Proof Intuition}
\label{sec-proof-intuition}

Our goal is to show that the skiplist template is linearizable. That is, we must prove that each of the core operations take effect in a single atomic step during its execution, the \emph{linearization point}, and satisfies the sequential specification shown in \cref{fig-search-str-spec}. For the skiplist template, we define the abstract state $\contentsof{N}$ to be the union of the \emph{logical contents} $\contentsof{n}$ of all nodes in $N$, where $\contentsof{n} \defeq \ite{\markof{n, 0}}{\emptyset}{\{\keyof{n}\}}$. In other words, the abstract state of the structure is a collection of keys contained in unmarked nodes at the base level.
\begin{figure}[h]
  \begin{align*}
  \dSpec{op}(k, C, C', \result)
  &\defeq
    \begin{cases}
      C' = C \land (\result \iff k \in C) & \cssOp = \code{search} \\
      C' = C \cup \set{k} \land (\result \iff k \not\in C) & \cssOp = \code{insert} \\
      C' = C \setminus \set{k} \land (\result \iff k \in C) & \cssOp = \code{delete}
    \end{cases}
  \end{align*}
  \caption{Sequential specification of a search structure. $k$ refers to the operation key, $C$ and $C'$ to the abstract state before and after operation $\cssOp$, respectively, and $\result$ is the return value of $\cssOp$.}
  \label{fig-search-str-spec}
\end{figure}
There are existing techniques from the literature that help us analyze the skiplist template. The two main techniques that we rely on are the \emph{Edgeset Framework}~\cite{DBLP:journals/tods/ShashaG88} and \emph{Hindsight Reasoning}~\cite{DBLP:conf/podc/OHearnRVYY10,DBLP:conf/wdag/Lev-AriCK15,DBLP:conf/wdag/FeldmanE0RS18,DBLP:journals/pacmpl/FeldmanKE0NRS20, DBLP:journals/pacmpl/MeyerWW22, DBLP:journals/pacmpl/0001W023}. We begin by giving a brief overview of the two techniques, proceeded by the analysis of the skiplist template using these techniques.

\subsection{The Edgeset Framework}
\label{sec-edgeset-framework}

The Edgeset Framework provides a common terminology to capture how search operations navigate  in a variety of search structures. We view each search structure as a mathematical graph whose edges are associated with an \emph{edgeset}, a label that is a set of keys. We denote the edgeset from $n$ to $n'$ by $\edgesetof{n,n'}$, and $k \in \edgesetof{n,n'}$ signifies that a search for key $k$  will proceed from node $n$ to $n'$. In the context of the skiplist template, we define the edgeset leaving $n$ to be all values greater than the key in $n$ if $n$ is unmarked. If node $n$ is marked, then the edgeset leaving $n$ is the entire keyspace. Formally: $\edgesetof{n,n'} \defeq \ite{n' = \nextof{n,0} \; \land \; \markof{n, 0} = \false}{(\keyof{n}, \infty)}{\KS}$. Note that, our definition of edgesets in the skiplist template  depends only on the base list, and not on higher level mark bits and next pointers. 

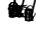
\begin{figure}[t]
\centering
  \begin{tikzpicture}[>=stealth, font=\footnotesize, scale=1, every node/.style={scale=0.8}]
  \def\xsep{1.5}
  \def\ysep{-1.8}

    \node[unode] (hd) {$0$};
    \node[stackVar, below=0.1mm of hd] (hd') {$\vhead$};

    \node[unode, right=6mm of hd] (n2) {$2$};
    \node[stackVar, below=0.1mm of n2] (n2') {$n_1$};

    \node[mnode, right=6mm of n2] (n4) {$4$};
    \node[stackVar, below=0.1mm of n4] (n4') {$n_2$};
    \node[mnode, right=6mm of n4] (n7) {$7$};
    \node[stackVar, below=0.1mm of n7] (n7') {$n_3$};
    \node[unode, right=6mm of n7] (n9) {$9$};
    \node[stackVar, below=0.1mm of n9] (n9') {$n_4$};
    \node[mnode, right=6mm of n9] (n15) {$15$};
    \node[stackVar, below=0.1mm of n15] (n15') {$n_5$};

    \node[unode, right=6mm of n15] (tl) {$\infty$};
    \node[stackVar, below=0.1mm of tl] (tl') {$\vtail$};

    \node[mnode, above=5mm of n9] (n9new) {$9$};


    \draw[edge] (hd) to (n2);
    \draw[edge] (n2) to (n4);
    \draw[edge] (n4) to (n7);
    \draw[edge] (n7) to (n9);
    \draw[edge] (n9) to (n15);
    \draw[edge] (n15) to (tl);
    \draw[edge] (n9new) to (n15);
    

\end{tikzpicture}
  \caption{Possible state of the base list in the skiplist template. Nodes are labeled with the value of their \code{key} field. Edges indicate \code{next} pointers. Marked (logically deleted) nodes are shaded gray. $\keysetof{\vhead} = \{0\}$, $\keysetof{n_1} = (0, 2]$, $\keysetof{n_4} = (2, 9]$ and $\keysetof{\vtail} = (9, \infty)$. The keyset of a marked node is always $\emptyset$.
  \label{fig:keyset-intuition}}
\end{figure}

A notion defined in terms of edgesets  is the \emph{inset} of a node, denoted by $\insetof{n}$, signifying a set of keys for which a search will arrive at $n$.  In order to understand the concept of inset intuitively,  consider \refFig{fig:keyset-intuition}. The inset of node $n_4$ is $(2, \infty)$, because, for all keys greater than 2, the search will enter $n_4$. We say node $n_1$ is the \emph{logical predecessor} of $n_4$ if it is the first unmarked predecessor of $n_4$.  The inset of the root is $\KS$ and the  inset of $n$ is the intersection of $\KS$ with the edgesets of all nodes between the root and $n$. For sorted linked lists in general,  a more local notion gives the same result: the inset of an unmarked node $n$ is $(\keyof{n'}, \infty)$, where $n'$ is the logical predecessor of $n$.

In contrast to inset, we define the \emph{outset} as the union of all its outgoing edgesets:  $\outsetof{n} \defeq \bigcup_{n' \in N} \edgesetof{n, n'}$.

We can now  define the \emph{keyset} of a node $n$ as $\keysetof{n} \defeq \insetof{n} \setminus \outsetof{n}$, i.e. intuitively, the set of keys for which a search enters $n$ but never leaves. The importance of keysets is that if $k$ is in $\keysetof{n}$, then $k$ is either in the contents of $n$ or is nowhere in the structure. In \refFig{fig:keyset-intuition}, the keyset of $n_4$ is $(2,9]$ and in general, the keyset of an unmarked node $n$ is $(\keysetof{n'}, \keyof{n}]$ where $n'$ is its logical predecessor. The keyset of a marked node is $\emptyset$ because its outset is the set of all keys $\KS$.

The technical definition of inset relies on the global data structure graph, defined as a solution to the following fixpoint equation
$$
  \forall n \in N.\; \inset(n) = \inflow(n) \cup \bigcup_{n' \in N} \edgeset(n', n) \cap \inset(n')
$$
where $\inflow(n) \defeq \ite{n = \vhead}{\KS}{\emptyset}$. Thus, the inset is a global quantity and hence difficult to reason about. Fortunately, this is where the Flow Framework~\cite{DBLP:journals/pacmpl/KrishnaSW18,DBLP:conf/esop/KrishnaSW20,DBLP:conf/tacas/MeyerWW23} comes in handy. It allows us to reason about quantities that can be expressed as a solution to a fixpoint equation (like inset) in a local manner by attaching \emph{flow} values to the node. The framework then provides tools to track changes to the flow values that are induced by changes to the underlying graph. Our approach to encoding keysets in Iris using the Flow Framework is borrowed from \cite{DBLP:conf/pldi/KrishnaPSW20}. We defer further details on this matter to the later sections.

As mentioned above, 
$\keysetof{n}$ intuitively is the set of all keys that $n$ is responsible for. Consider \Cref{fig:keyset-intuition} again, a thread executing \code{search(6)} without any interference will reach node $n_4$ and terminate, concluding that $6$ is not present in the structure. In this sense, we say $n_4$ is responsible for key $6$ and therefore $6$ is part of $n_4$'s keyset. The keysets of all nodes partition the set of all keys and provide the crucial \emph{Keyset Property}:
\begin{equation}
  \label{eqn-keyset}
    \forall \; n \in N, k \in \keyspace. \; k \in \keysetof{\anode} \Rightarrow (k \in \contentsof{N} \Leftrightarrow k \in \contentsof{\anode})
    \tag{\textsf{KeysetPr}}
  \end{equation}

This property enables one to lift a proof of the specification at the node level to a proof of the sequential specification $\dSpec{op}$. A particular situation where ($\ref{eqn-keyset}$) proves indispensable is when \code{search} fails to find the search key. Note that \code{search}  observes only the nodes it visited, and hence  has only a partial view of the structure. When \code{search} fails to find the key, the proof has to reconcile this partial view of the structure with the global view. In essence, if a concurrent invocation of \code{search} on key $k$ fails to find the key, can we conclude that there was a point in time during its execution when $k$ was in fact not present in the structure? Here, the property ($\ref{eqn-keyset}$) helps us reconcile facts gathered by search with the global state of the structure. Specifically, if \code{search} can determine a node $n$ such that $k \in \keysetof{n}$ and $k \notin \contentsof{n}$, then we can immediately infer that $k$ was not present in the structure at that point in time.

\subsection{Hindsight Reasoning}
\label{sec-hindsight-intro}


Lock-free structures often exhibit future-dependent linearization points. That is, the linearization point of an operation cannot be determined at any fixed moment, but only at the end of the execution, once any interference of other concurrent operations has been accounted for. To understand the interference issue, consider the \code{search} operation. Since, \code{search}  returns the result of \code{traverse}, let us look at the eager traversal implementation. To simplify the explanation further, let us assume that the maximum height allowed for every \np{non-sentinel} node is one. Then, we can ignore the $\eagerRec$ function and focus on $\eagerI$ called at the base level.   

Let there be a thread $T$ executing \code{search($7$)}. Concurrently, there is a thread $T_d$ executing \code{delete($7$)} and a thread $T_i$ executing \code{insert($7$)}. \Cref{fig:traversal-scenarios} shows interesting scenarios that thread $T$ might potentially observe. Box (a) captures the state of the structure at the beginning of the $\eagerI$ call processing $n_2$. Let \textbf{Scenario~1} be the situation when thread $T$ faces no interference from $T_d$ and $T_i$. Here, thread $T$ finds the key $7$ in $n_2$ and $\eagerI$ returns $\true$. The point when $\eagerI$ finds $n_2$ to be unmarked becomes the linearization point for this scenario.

\begin{figure}[t]
\centering
  \begin{tikzpicture}[>=stealth, font=\footnotesize, scale=1, every node/.style={scale=0.8}]
  \def\xsep{1.5}
  \def\ysep{-1.8}

  \begin{scope}[local bounding box=g1]
    
    \node[stackVar] (inf) {};

    \node[unode, right=6mm of inf] (n4) {$4$};
    \node[stackVar, above=0.1mm of n4] (n4') {$n_1$};
    \node[unode, right=6mm of n4] (n7) {$7$};
    \node[stackVar, above=0.1mm of n7] (n7') {$n_2$};
    \node[unode, right=6mm of n7] (n9) {$9$};
    \node[stackVar, above=0.1mm of n9] (n9') {$n_3$};

    \node[stackVar, right=6mm of n9] (Inf) {};

    \node[unode, white, above=5mm of n7] (n7new) {$7$};

    \node[stackVar, below=.4cm of n4] (pred) {$\vpred$};
    \node[stackVar, below=.4cm of n7] (curr) {$\vcurr$};

    \draw[edge] (pred) to (n4);
    \draw[edge] (curr) to (n7);
    \draw[dashed, ->] (inf) to (n4);
    \draw[edge] (n4) to (n7);
    \draw[edge] (n7) to (n9);
    \draw[dashed, ->] (n9) to (Inf);
    
    \draw ($(n4.north west)+(-.9,1.4)$) rectangle ($(n9.south east)+(.9,-1)$);
    \node at ($(n4.north west)+(-.7, 1.2)$) {$(a)$};

  \end{scope}

  \begin{scope}[local bounding box=g2, shift={($(g1.east)+(2, -.2)$)}]
    \node[stackVar] (inf) {};

    \node[unode, right=6mm of inf] (n4) {$4$};
    \node[stackVar, above=0.1mm of n4] (n4') {$n_1$};
    \node[mnode, right=6mm of n4] (n7) {$7$};
    \node[stackVar, above=0.1mm of n7] (n7') {$n_2$};
    \node[unode, right=6mm of n7] (n9) {$9$};
    \node[stackVar, above=0.1mm of n9] (n9') {$n_3$};

    \node[stackVar, right=6mm of n9] (Inf) {};

    \node[unode, white, above=5mm of n7] (n7new) {$7$};

    \node[stackVar, below=.4cm of n4] (pred) {$\vpred$};
    \node[stackVar, below=.4cm of n7] (curr) {$\vcurr$};

    \draw[edge] (pred) to (n4);
    \draw[edge] (curr) to (n7);
    \draw[dashed, ->] (inf) to (n4);
    \draw[edge] (n4) to (n7);
    \draw[edge] (n7) to (n9);
    \draw[dashed, ->] (n9) to (Inf);
    
    \draw ($(n4.north west)+(-.9,1.4)$) rectangle ($(n9.south east)+(.9,-1)$);
    \node at ($(n4.north west)+(-.7, 1.2)$) {$(b)$};

  \end{scope}

  \begin{scope}[local bounding box=g3, shift={($(g2.south)+(-2.4,-3.0)$)}]
    \node[stackVar] (inf) {};

    \node[unode, right=6mm of inf] (n4) {$4$};
    \node[stackVar, above=0.1mm of n4] (n4') {$n_1$};
    \node[mnode, right=6mm of n4] (n7) {$7$};
    \node[stackVar, above=0.1mm of n7] (n7') {$n_2$};
    \node[unode, right=6mm of n7] (n9) {$9$};
    \node[stackVar, above=0.1mm of n9] (n9') {$n_3$};

    \node[stackVar, right=6mm of n9] (Inf) {};

    \node[unode, white, above=5mm of n7] (n7new) {$7$};

    \node[stackVar, below=.4cm of n4] (pred) {$\vpred$};
    \node[stackVar, below=.4cm of n9] (curr) {$\vcurr$};

    \draw[edge] (pred) to (n4);
    \draw[edge] (curr) to (n9);
    \draw[dashed, ->] (inf) to (n4);
    \draw[edge] (n4) to[out=60, in=120] (n9);
    \draw[edge] (n7) to (n9);
    \draw[dashed, ->] (n9) to (Inf);
    
    \draw ($(n4.north west)+(-.9,1.4)$) rectangle ($(n9.south east)+(.9,-1)$);
    \node at ($(n4.north west)+(-.7, 1.2)$) {$(c)$};

  \end{scope}

  \begin{scope}[local bounding box=g4, shift={($(g1.south)+(-2.4,-3.0)$)}]
    \node[stackVar] (inf) {};
    \node[unode, right=6mm of inf] (n4) {$4$};
    \node[stackVar, above=0.1mm of n4] (n4') {$n_1$};
    \node[mnode, right=6mm of n4] (n7) {$7$};
    \node[stackVar, above=0.1mm of n7] (n7') {$n_2$};
    \node[unode, right=6mm of n7] (n9) {$9$};
    \node[stackVar, above=0.1mm of n9] (n9') {$n_3$};
    \node[unode, above=5mm of n7] (n7new) {$7$};
    \node[stackVar, right=1mm of n7new] (n7new') {$n_4$};
    \node[stackVar, right=6mm of n9] (Inf) {};

    \node[stackVar, below=.4cm of n4] (pred) {$\vpred$};
    \node[stackVar, below=.4cm of n9] (curr) {$\vcurr$};

    \draw[edge] (pred) to (n4);
    \draw[dashed, ->] (inf) to (n4);
    \draw[edge] (n7) to (n9);
    \draw[dashed, ->] (n9) to (Inf);
    
    \draw[edge] (n4) to[bend left=20] (n7new);
    \draw[edge] (n7new) to[bend left=20] (n9);
    \draw[edge] (curr) to (n9);
    \draw ($(n4.north west)+(-.9,1.4)$) rectangle ($(n9.south east)+(.9,-1)$);
    \node at ($(n4.north west)+(-.7, 1.2)$) {$(d)$};

  \end{scope}


  \draw[edge,dotted] ($(g1.east)+(0,0)$) to node[above] {\large\begin{tabular}{l}\code{delete($7$)}\\\;\rotatebox[origin=c]{270}{$\rightsquigarrow$}\end{tabular}} ($(g2.west)+(0,0)$);

  \draw[edge,dotted] ($(g2.south)+(0,0)$) to ($(g3.north)+(0,0)$);

  \draw[edge,dotted] ($(g3.west)+(0,0)$) to node[below] {\large\begin{tabular}{l}\;\rotatebox[origin=c]{90}{$\rightsquigarrow$} \\ \code{insert($7$)} \end{tabular}}($(g4.east)+(0,0)$);


\end{tikzpicture}
  \caption{Possible states of \code{search($7$)} on the base level in presence of interference from concurrent \code{delete($7$) and \code{insert($7$)}}.
  \label{fig:traversal-scenarios}}
\end{figure}

Now consider \textbf{Scenario~2} to be the situation where thread $T_d$ marks $n_2$ before $\eagerI$ processes it, as shown in Box (b). Thread $T$ will attempt to unlink $n_2$, and assuming no further interference, the unlink will result in the structure in Box (c). Thread $T$ will process $n_3$ next, finding $n_3$ to be unmarked with key greater than $7$, and will  terminate with result $\false$. So when is the linearization point in this scenario? It cannot be when $T$ finds $n_3$ unmarked when processing it. Because there could be further interference from thread $T_i$ which inserts key $7$ in a new node as shown in Box (d). The new node could be added right before $T$ reads the mark bit of $n_3$. Thus, when $\eagerI$ finds $n_3$ unmarked and returns $\false$, key $7$ could actually be present in the structure at that point in time.

The linearization point is actually the point in time shown in Box (c), i.e., right after $n_2$ is unlinked. However, thread $T$ cannot confirm this when $n_2$ is unlinked because $\eagerI$ may not terminate at $n_3$ with $\false$ as the result. The reason is that by the time $T$ processes $n_3$, it could get marked in a manner similar to $n_2$ in Box (b), resulting in the unlinking of $n_3$ and potentially a restart. That Box (c) is the linearization point is confirmed when $T$ has found $n_3$ to be unmarked later. The structure maintains the invariant that once a node is marked, it remains marked. Using this invariant, an analysis of thread $T$'s history concludes that $n_3$ must have been unmarked at the  point when $n_2$ was unlinked. Once $\eagerI$ terminates at $n_3$ with $\false$, an analysis can \emph{establish in hindsight} that Box (c) indeed was the linearization point. 

\newcommand{\vprop}{\mathit{q}}
\newcommand{\vpropp}{\mathit{r}}

Hindsight reasoning as formalized
in~\cite{DBLP:journals/pacmpl/MeyerWW22,DBLP:journals/pacmpl/0001W023}
is designed to deal with situations like the \code{search} in
\cref{fig:traversal-scenarios}. It enables temporal reasoning about
computations using a \emph{past predicate} $\past \vprop$, which
expresses that proposition $\vprop$ held true at some prior state in
the computation (up to the current state). For instance,
$\past (\nextof{n_1, 0} = n_2)$ holds in Box (c) even though
$\nextof{n_1, 0} = n_3$ at that point. The reason is that
$\nextof{n_1, 0} = n_2$ was true at an earlier point in time, namely
in Box (b). Note that the past operator $\past$ abstracts away the
exact time point when the predicate held true. Note also that a past
predicate is not affected by concurrent interferences, as it merely
records some fact about a past state.

There are two ways to establish a past predicate that are relevant for our proofs. The first is to establish the predicate in the current state directly. That is, $\past \vprop$ holds if $\vprop$ holds in the current state. As an example, we obtain $(\nextof{n_1, 0} = n_2)$ when \code{findNext} on $n_1$ returned $n_2$ in Box (a). Thus, for all subsequent states including Box (b) and (c), we get $\past (\nextof{n_1, 0} = n_2)$. The second way to establish a past predicate is through the use of \emph{temporal interpolation}~\cite{DBLP:journals/pacmpl/0001W023}. That is, one proves a lemma of the form: if there existed a past state that satisfied property $\vprop$ and the current state satisfies $\vpropp$, then there must have existed an intermediate state that satisfied $o$. Such lemmas can then be applied, e.g., to prove that if thread $T$ finds $n_3$ to be unmarked in Scenario~2, then it must have been unmarked when $n_2$ was unlinked in Box (c).

Equipped with the Edgeset Framework and hindsight reasoning, we are now ready to analyze the core operations of the skiplist template.  

\subsection{Proof Outline for Core Operations}
\label{sec-proof-outline}

We refer to a linearization point as \emph{modifying} if the operation changes the abstract state of the data structure (like in the case of a succeeding \code{delete} and \code{insert}) and otherwise refer to it as \emph{unmodifying} (like \code{search} and in the case of a failing \code{delete} or \code{insert}). The modifying linearization points of the skiplist template are easier to reason about because they are not future-dependent. For \code{delete}, the linearization point occurs when \code{markNode} succeeds, and similarly, for \code{insert}  the linearization point occurs when the call to \code{changeNext} on line~\ref{line-insert-changeNext} succeeds. 
The proof strategy for unmodifying linearization points  is to combine (\ref{eqn-keyset}) with the $\past$ operator from hindsight reasoning. Let us expand on this proof strategy in detail and show why the skiplist template is linearizable.

We begin by describing the specification for \code{traverse} that is assumed for analyzing the core operations of the template. Then, we analyze each of the operations in detail. Finally, we show how the eager implementations of \code{traverse} satisfies the specification that was assumed in the beginning. Along the way, we introduce (as and when necessary) invariants maintained by the skiplist template that are crucial for proving linearizability.

\textbf{Specification of \code{traverse}.} The function $\code{traverse}\,\vparr\,\vsarr\,k$ updates arrays $\vparr$ and $\vsarr$ with predecessor-successor pairs for each level and returns a triple $(\vpred, \vcurr, \vres)$ that satisfies the following past predicate regarding node $\vcurr$: $\past{(k \in \keysetof{\vcurr} \land ( \vres \iff k \in \contentsof{\vcurr}))}$. Recall that  our definition of edgesets in \cref{sec-edgeset-framework} implies the following invariant: 

\makeatletter
\let\orgdescriptionlabel\descriptionlabel
\renewcommand*{\descriptionlabel}[1]{%
  \let\orglabel\label
  \let\label\@gobble
  \phantomsection
  \edef\@currentlabel{#1\unskip}%
  \let\label\orglabel
  \orgdescriptionlabel{#1}%
}
\makeatother

\begin{enumerate}[label=\textbf{Invariant~\arabic*},ref={\arabic*},itemindent=60pt]
\item \label{inv-sk1} For all nodes $n$, if $\markof{n, 0}$ is set to $\true$ then $\keysetof{n} = \emptyset$. 
\end{enumerate}
Using Invariant~\ref{inv-sk1}, we can establish that $\vcurr$ is unmarked at the base level at the time point when $k \in \keysetof{\vcurr}$ holds. \np{Note that \code{traverse} may physically unlink marked nodes. However, this step does not change the abstract state of the structure. Hence, the specification for \code{traverse} involves no change of the abstract state.}

We now consider each of the core operations in detail.

\textbf{Proof of \code{search}.} 
Function \code{search} returns $\vres$ out of the triple $(\vpred, \vcurr, \vres)$ returned by \code{traverse}. 
The specification of \code{traverse} says $\vres \iff k \in \contentsof{\vcurr}$ at some point, say $t$, during its execution. The specification additionally guarantees $k \in \keysetof{\vcurr}$ at time $t$. These two facts, combined with the (\ref{eqn-keyset}) at time point $t$, allow us to immediately infer that $\vres$ is true iff $k$ was in the structure at that point.
Hence, we can establish that $(\vres \iff k \in \contentsof{\vcurr})$ was true at some point during the execution of \code{search}.

\textbf{Proof of \code{delete}.} We analyze \code{delete} by case analysis on the value $\vres$ returned by \code{traverse}. If $\vres$ is $\false$, then again we can establish that $k$ was not in the structure at some point during \code{traverse}'s execution by the same reasoning used in the proof of \code{search}. So let us consider the case that $\vres$ is $\true$. By the specification of $\code{traverse}$, we can establish a time point when $\vcurr$ was unmarked and contained $k$. The \code{delete} operation then calls \code{maintainanceOp\_del} which marks $\vcurr$ at all the higher levels. Finally, the \code{markNode} on Line~\ref{line-delete-markNode} attempts to mark $\vcurr$ at the base level. If \code{markNode} succeeds, then this step becomes the linearization point of \code{delete} and $k$ can be considered to be deleted from the structure. But if \code{markNode} fails, then we gain the knowledge that $\markof{\vcurr, 0} = \true$. Hindsight reasoning allows us to infer that $\vcurr$ was marked at the base level by a concurrent thread between the end of \code{traverse} and the invocation of \code{markNode}. The point right after $\vcurr$ was marked by a concurrent thread becomes the linearization point of \code{delete} in this case, as we can determine that $k$ was not present in the structure at that point.

This hindsight reasoning relies on two facts: first, the key of a node never changes and second, once a mark bit is set to $\true$ by a successful \code{markNode} operation (at line~\ref{line-delete-markNode} in \code{delete} or line~\ref{line-delete-mnt-markNode} in \code{maintainanceOp\_del}), no other operation will set it back to $\false$. In fact, these two facts are invariant for the skiplist template:
\begin{enumerate}[resume*]
\item \label{inv-sk2} For all nodes $n$ and level $i$, once $\markof{n, i}$ is set to $\true$, it remains $\true$. 
\item \label{inv-sk3} For all nodes $n$, $\keyof{n}$ remains constant. 
\end{enumerate}

\textbf{Proof of \code{insert}.} Similar to \code{delete}, we begin by case analysis on $\vres$ returned by \code{traverse}. If $\vres$ is true, then we can establish that $k$ was already present in the structure at some point. Otherwise, $\vres$ is $\false$ and \code{insert} creates a new node $\ventry$ with key $k$. Using \code{changeNext}, an attempt is made to insert node $\ventry$ between nodes $\vpred$ and $\vcurr$. If the attempt succeeds, then $k$ is now part of the structure and this becomes the linearization point. The following \code{maintainanceOp\_ins} operation does not change the abstract state of the structure, and thus, has no effect in terms of linearizability. If the \code{changeNext} fails, then \code{insert} simply restarts.

As is evident with the proof outline for the core operations, the specification assumed for \code{traverse} plays a critical role in case the operation exhibits an unmodifying linearization point. Let us now turn to \code{traverse} and show how its specification can be proved. We analyze the eager traversal in detail in the following section. The proof argument for the lazy version is similar. 

\subsection{Proof Outline for Eager Traversal}

As stated earlier, \code{traverse} returns $(\vpred, \vcurr, \vres)$ such that $\past{(k \in \keysetof{\vcurr} \land (\vres \iff k \in \contentsof{\vcurr}))}$. Since the returned triple is the result of a call to $\eagerI$ at the base level, let us begin by analyzing the behavior of this call.

In the sequential setting, the traversal in a search structure maintains the invariant that the search key is always in the inset of the current node. This invariant holds by the design of the Edgeset Framework. Unfortunately, this invariant no longer holds for the skiplist template in the concurrent setting as evidenced by Box (c) in \cref{fig:traversal-scenarios}. However, we argue first that $\eagerI$ does maintain the invariant that the search key was in the inset of the current node $\vcurr$ between the start of the traversal and the point at which the $\eagerI$ accesses $\vcurr$. We call this the \emph{inset in hindsight} invariant.

We prove this invariant inductively. We make use of the following locally maintained invariants: (i) At all times, there is one list, denoted the \emph{reachable list}, from the head node that includes all unmarked and some marked nodes. (This list is characterized by the set of nodes with non-empty inset, see \cref{fig:keyset-intuition} for intuition).  (ii) The keys in the reachable list are sorted. A consequence of these two invariants is that if a node $n$ is in the reachable list (whether $n$ is marked or not) and has a key less than $k$, then $k$ is in the inset of $n$.

To prove that inset in hindsight is an invariant, we have to show that (a) it is an invariant when $\eagerI$ takes a step (\cref{eager-tri-findNext}) when traversing the base level, and (b) that we can establish inset in hindsight when $\eagerRec$ initiates $\eagerI$ (\cref{eager-rec-eageri}) at the base level.

To show (a), observe that if a node $n$ becomes unlinked from the reachable list, then it will never again be part of the reachable list. Hence, if $n$ is not in the reachable list when $\eagerI$ begins executing at the base list, then $\eagerI$ will never visit $n$. The contrapositive of this statement allows us to say that if $\eagerI$ reaches some node $\vcurr$, then it must have been part of the reachable list at some point during the execution of $\eagerI$. Additionally, $\eagerI$ proceeds to the node following $\vcurr$  only when $\keyof{\vcurr} < k$. With the help of invariants (i) and (ii) above, we can thus establish that $k$ was in the inset of $n$ at some point.

To show (b), we must do a case analysis on whether node $\vpred$ (\cref{eager-rec-findNext}) is marked. If it is unmarked, then it is straightforward to establish that $k$ is in the inset of $\vcurr$ currently. However, if $\vpred$ is marked, then we require temporal interpolation based on the following invariant:
\begin{enumerate}[resume*]
  \item \label{inv-sk6}  For all nodes $n$ and level $i$, once $\markof{n, i}$ is set to $\true$, $\nextof{n, i}$ does not change.
\end{enumerate}
This invariant tells us that if $\vpred$ was known to be unmarked in the past, and it is marked currently, then $p$ must have been pointing to $\vcurr$ right before it got marked. At that point in time, we can establish that $k$ must have been in the inset of $\vcurr$. 

This completes the inductive proof that inset in hindsight is indeed an invariant maintained by the traversal. The inset in hindsight invariant is sufficient to prove the \code{traverse} specification by the following simple argument.  If the traverse encounters $k$ in an unmarked node $n$, then \code{traverse} will return $\true$ as it should. If, by contrast, \code{traverse} encounters an unmarked node $n$ such that $\keyof{n} > k$, then by the inset in hindsight invariant, $k$ must have been in the inset of $n$ at some point $t$ in the past and $k$ cannot be in the outset of $n$ (because $\keyof{n} > k$ and $n$ is unmarked), so therefore $k$ must have been in the keyset of $n$ at time $t$.



\section{Hindsight Reasoning in Iris}
\label{sec-hindsight}

\newcommand{\minvr}{\sinv{\mainInv(r)}}

Linearizability in Iris is defined via \emph{(logically) atomic triples}~\cite{DBLP:conf/ecoop/PintoDG14, DBLP:conf/popl/JungSSSTBD15}. Intuitively, an atomic triple $\atomicTriple{x.\;P}{e}{\Ret\val. Q}$ says that at some point during the execution of $e$, the resources described by the precondition $P$ will be updated to satisfy the postcondition $Q$ for return value $\Ret\val$ in one atomic step. The variable $x$ can be thought of as the abstract state of the data structure before the update at the linearization point.

Linearizability of a search structure operation $\cssOp$ can be expressed by an atomic triple of the form
\begin{equation}
  \minvr \magicwand \atomicTriple{C.\; \cssstate(r, C)}{\cssOp\,r\,k}{\vres.\; \exists\; C'. \; \cssstate(r, C') \,*\, \dSpec{op}(k, C, C', res) }. \tag{\textsf{ClientSpec}}\label{eq:client-spec}
\end{equation}
Here, $r$ is the pointer to the head of the data structure. The predicate $\cssstate(r, C)$ is the \emph{representation predicate} that relates the head pointer with the contents $C$ of the structure. The predicate $\mainInv(r)$ is the shared data structure invariant. It can be thought of as a thread-local precondition of the atomic triple, which we express using separating implication. The invariant ties $\cssstate(r, C)$ to the data structure's physical representation and may contain other resources necessary for proving the atomic triple. The predicate $\dSpec{op}(k, C, C', res)$ captures the sequential specification of the structure. The specification essentially says there is a single atomic step in $\cssOp$ where the abstract state changes from $C$ to $C'$ according to the sequential specification  $\dSpec{op}(k, C, C', res)$ (\cref{fig-search-str-spec}). This step is $\cssOp$'s linearization point. We call (\ref{eq:client-spec}) the \emph{client-level} atomic specification for the data structure under proof.

\smartparagraph{Proving atomic triples}
The proof of establishing an atomic triple involves a \emph{linearizability obligation} that must be discharged directly at the linearization point. However, it can be challenging to determine the linearization point precisely and to discharge the linearizability obligation exactly at that point.
When the program execution reaches a potential linearization point that depends on future interferences by other threads, then the proof will fail if it is unable to determine whether the linearizability obligation should be discharged now or later. In Iris, this challenge is overcome using \emph{prophecy variables}~\cite{DBLP:journals/pacmpl/JungLPRTDJ20}, which enable the proof to reason about the remainder of the computation that has not yet been executed.

Another challenge is that the linearization point of an operation may be an atomic step of another operation that is executed by a different thread (like in Scenario~2 discussed in \cref{sec-hindsight-intro}). Data structures that demonstrate such behavior are said to deploy \emph{helping}. This behavior complicates thread modular reasoning. The conventional solution to this challenge in Iris is to use a \emph{helping protocol}~\cite{DBLP:journals/pacmpl/JungLPRTDJ20,DBLP:journals/pacmpl/PatelKSW21,DBLP:journals/pacmpl/JungLCKPK23}. The helping protocol is specified as part of the shared data structure invariant and consists of a registry that tracks which threads are expected to be linearized by other threads as well as conditional logic that governs the correct transfer and discharge of the associated linearizability obligations.

Both the use of prophecy variables and the helping protocol need to be tailored to the specific data structure at hand, which adds considerable overhead to the proof. To reduce this overhead, we present an alternative proof method that enables linearizability proofs based on hindsight arguments in Iris. Rather than identifying the linearization point precisely, the proof can establish linearizability in hindsight using temporal interpolation in the style of the intuitive proof argument for the skiplist template presented in \cref{sec-hindsight-intro}.

\smartparagraph{Hindsight specification}
Our proof method offers an intermediate specification, a Hoare triple specification, which in essence expresses that linearizability has been established in hindsight. 
%
In our Iris formalization, we show that any data structure whose operations satisfy the hindsight specification also satisfy the client-level atomic specification. This proof relates the two specifications via prophecy variables and a helping protocol. However, the helping protocol is data structure agnostic, making our proof method applicable to a broad class of structures exhibiting future-dependent unmodifying linearization points.

From the perspective of a proof author using our method to prove linearizability of some structure, one has to only establish the hindsight specification to obtain the proof of the client-level atomic specification. To this end, our method provides further guidance to the proof author.

In order to use hindsight reasoning, one has to have the history of computation at hand. Here, we offer a shared state invariant with a mechanism to store the history. The shared state invariant has three main components: a mechanism to store the history, the helping protocol, and finally, an abstract predicate that can be instantiated with invariants specific to the structure at hand. The first two components are data structure agnostic. The proof author only needs to specify the data structure-specific invariant and what information about the data structure state should be tracked by the history.


In the rest of this section, we discuss our method in detail. We begin with the hindsight specification, followed by a discussion of the shared state invariant and how to use it. 

\subsection{Linearizability in Hindsight}
\label{sec-hindsight-spec}

\newcommand{\csslop}{\overline{\cssOp}}
\newcommand{\vpvs}{\mathit{pvs}}
\newcommand{\vpvss}{\mathit{pvs'}}
\newcommand{\vprf}{\mathit{prf}}
\newcommand{\prophend}[1]{\code{END}(#1)}
\newcommand{\procProph}{\mathsf{Upd}}
\newcommand{\valProph}{\mathsf{Val}}

We motivate the hindsight specification using the challenges we face when proving the client-level atomic specification for the \code{delete} operation of the skiplist template. 
Let us recall the intuitive proof argument for \code{delete} from \Cref{sec-proof-outline}. As per the observation regarding the modifying and unmodifying linearization points, a \code{delete} thread with modifying linearization point can fulfill the obligation at the point when the structure is modified. However, a \code{delete} thread with an unmodifying linearization point requires helping. 

\smartparagraph{Prophecy reasoning}
An important detail of our proof method is how it determines whether a thread requires helping. In the following, we refer to the operation that a thread performs at its linearization point as its \emph{decisive operation}. In \code{delete}, the traversal observes node $\vcurr$ to be unmarked at some point during execution. In the case where $\vcurr$ is marked by the time that the thread calls its decisive operation $\code{markNode}$ (in Line~\ref{line-delete-markNode}), the thread requires helping from the thread that marks $\vcurr$. 

In order to determine in advance whether a thread requires helping, our proof method attaches a prophecy to each thread. A prophecy in Iris can predict a sequence of values and is treated as a resource that can be owned by a thread. Ownership of a prophecy $p$ is captured by the predicate $\proph(p, \vpvs)$, where $\vpvs$ is the list of predicted values. The predicate signifies the right to resolve $p$ when the thread makes a physical step that produces some result value $v$. The resolution of $p$ establishes equality between $v$ and the head of the list $\vpvs$ (i.e., the next value predicted by $p$). The resolution step yields the updated predicate $\proph(p, \vpvs')$ where $\vpvs'$ is the tail of $\vpvs$.
This mechanism enables the proof to do a case analysis on the predicted values $\vpvs$ before these values have been observed in the program execution\footnote{For further details on prophecies in Iris, we refer to \cite{DBLP:journals/pacmpl/JungLPRTDJ20}.}. 

The prophecy attached to a thread predicts the results of the thread's decisive operation. In case of \code{delete}, the decisive operation is the call to \code{markNode} in the base list, while for \code{insert}, it is the call to \code{changeNext} in the base list. Note that a thread may restart if its decisive operation fails (like in the case of \code{insert}). Therefore, the prophecy needs to predict a sequence of result values, one for each attempted call to the thread's decisive operation.

For the purpose of this discussion, we assume that the prophecy predicts a sequence of \code{Success} or \code{Failure} values. If the sequence contains a \code{Success} value, then the decisive operation will succeed and the thread will modify the structure. Otherwise, the thread's linearization point is unmodifying.
Let predicate $\procProph(\vpvs)$ hold when $\vpvs$ contains at least one \code{Success} value.

The proof author only needs to identify the decisive operations that potentially change the abstract state of the structure (like \code{markNode} as discussed above) by resolving the prophecy around these decisive calls.


\smartparagraph{Hindsight specification}
Before we can present the hindsight specification, we need to provide necessary details regarding the atomic triples in Iris. An atomic triple $\atomicTriple{x.\;P}{e}{\Ret\val. Q}$ is defined in terms of standard Hoare triples of the form $\All\; \Phi. \hoareTriple{\atomicUpdate_{x. P, Q}(\Phi)}{e}{\Ret\val. \Phi(\val)}$. The predicate $\atomicUpdate_{x. P, Q}(\Phi)$ is the \emph{atomic update token} and represents the linearizability obligation of the atomic triple. At the beginning of each atomic step that the thread takes up to its linearization point, the token offers the resources in $P$ and the token itself transforms into a choice. That is, at the end of the atomic step, the prover has to chose to either \emph{commit} the linearization or \emph{abort}. When committing, the prover has to show that the thread's atomic step transforms the resources in $P$ to those in $Q$, receiving $\Phi(v)$ from the update token in return, which serves as the receipt of linearization of the atomic triple. In case of an abort, the prover needs to show that the thread's atomic step reestablishes $P$.

We also need to introduce two more auxiliary predicates:

\newcommand{\vtid}{\mathit{tid}}

\begin{itemize}
  \item $\thread(\vtid, t_0)$: this predicate is used to \emph{register} the thread with identifier $\vtid$ in the shared invariant. The argument $t_0$ denotes the time when thread $\vtid$ began its execution.
  \item $\pastlin(\cssOp, k, \vres, t_0)$: this predicate holds if there was a past state in the history between time $t_0$ and the point when this predicate is evaluated for which the sequential specification $\dSpec{op}$ held with result $\vres$. It essentially captures whether the sequential specification was true for any point after time $t_0$.
\end{itemize}


\newcommand{\pcolor}{brown}
\newcommand{\ucolor}{teal}
\newcommand{\ncolor}{magenta}

We now have all the ingredients to present the hindsight specification:
\begin{equation}
  \label{eq:hindsight-spec}
  \begin{array}{l}
  \forall \; \tid \; t_0 \; \mathit{pvs}.\; \minvr \magicwand \thread(\tid, t_0) \magicwand\\
  \annot{\textcolor{\pcolor}{\proph(p, \vpvs)} \;*\; \textcolor{\ucolor}{(\procProph(pvs) \magicwand \atomicUpdate_{\cssOp}(\Phi))}} \; \cssOp\;r\;k \\
  \annot{\vres.\; \textcolor{\pcolor}{\exists \vpvss. \; \proph(p, \vpvss) \;*\; \vpvs = (\_ \;\code{@}\; \vpvss)} \\ \;*\; \textcolor{\ucolor}{(\procProph(pvs) \magicwand \Phi(\vres))} \\ \;*\; \textcolor{\ncolor}{(\neg \procProph(pvs) \magicwand \pastlin(\cssOp, k, \vres, t_0))}} \tag{\textsf{HindSpec}}
\end{array}
\end{equation}
We explain it piece by piece. The local precondition $\thread(\tid, t_0)$ ties the thread to its identifier $\tid$ and provides knowledge that $\tid$ begins executing at time $t_0$. The Hoare triple can be best understood by observing how prophecy resources are allowed to change (highlighted in \textcolor{\pcolor}{\pcolor}) and what are the obligations when $\procProph(\vpvs)$ holds (in \textcolor{\ucolor}{\ucolor}) versus when it does not hold (in \textcolor{\ncolor}{\ncolor}). Let us look at each of these in detail. First, the prophecy resource $\proph(p, \vpvs)$ in the precondition changes to $\proph(p, \vpvss)$ in the postcondition where $\vpvss$ is a suffix of $\vpvs$. It basically says that operation $\cssOp$ is allowed to resolve the prophecy $p$ as many times as it needs and then return the remaining resource at the end. 

Now let us consider the case when $\procProph(\vpvs)$ holds. The precondition here provides the atomic update token $\atomicUpdate_{\cssOp}(\Phi)$ to $\cssOp$, expecting the receipt of linearization $\Phi(\vres)$ in return. Thus, the responsibility of linearization is delegated to $\cssOp$ when $\procProph(\vpvs)$ holds. We can gain better insight by relating this situation to the \code{delete} operation from the skiplist template as before. This case corresponds to when \code{markNode} (from line~\ref{line-delete-markNode}) succeeds as $\procProph(\vpvs)$ holds here. The point when \code{markNode} succeeds becomes the linearization point and so the thread does not require help from other threads to linearize. The hindsight specification simply asks for the receipt from linearization $\Phi(res)$ at the end. 

Finally, let us consider the case when $\procProph(\vpvs)$ does not hold. The precondition provides no additional resources here, while the postcondition requires the predicate $\pastlin(\cssOp, k, \vres, t_0)$. In simple terms, this means that if $\procProph(\vpvs)$ is not true, i.e., the prophecy says the thread is not going to modify the structure, then the hindsight specification allows exhibiting a past state from history when the sequential specification was true. Relating again to \code{delete}, if the \code{markNode} fails, then the thread can look at the history of the structure and exhibit precisely the point when the decisive node got marked.

The proof argument for establishing the hindsight specification is significantly simpler than if one were to attempt a direct proof of the client-level atomic specification. In particular, the proof author does not need to reason about helping and atomic update tokens in last case discussed above. Instead, they only need to reason about the structure-specific history invariant.

\smartparagraph{Soundness of the hindsight specification}
Our proof that relates the hindsight specification for $\cssOp$ to the atomic triple specification involves a helping protocol. \np{The details of the helping protocol and the soundness proof for the hindsight specification are similar to those of the proofs presented in~\cite{DBLP:journals/pacmpl/JungLPRTDJ20,DBLP:journals/pacmpl/PatelKSW21}. We therefore  provide only a brief summary here. Additional details regarding the proof and the helping protocol can be found in \cref{sec-hindsight-proof}.}

Before $\cssOp$ begins executing, the proof creates the prophecy resource $\proph(p, \vpvs)$ assumed in the precondition of the hindsight specification. If the prophecy determines that the thread requires helping, then its client-level atomic triple is registered to a predicate which encodes the helping protocol as part of the shared state invariant $\mainInv(r)$. The registered atomic triple serves as an obligation for the helping thread to commit the atomic triple. This obligation will be discharged by the appropriate concurrent operation determined by the $\cssOp$'s sequential specification $\dSpec{op}$. The proof then uses the hindsight specification to conclude that it can collect the committed triple from the shared predicate. \np{The committed triple serves as a receipt that the obligation to linearize has been fulfilled.}

\np{
 To govern the transfer of linearizability obligations and fulfillment receipts between threads via the shared invariant, the helping protocol tracks a \emph{registry} of thread IDs with unmodifying linearization points that require helping from other concurrent threads. Each thread registered for helping is in either \emph{pending} state or \emph{done} state, depending on whether the thread has already been linearized. A thread registered for helping must be able to determine its current protocol state in order to be able to extract its committed atomic triple from the registry. For this purpose, the helping protocol includes a \emph{linearization condition} that holds iff a registered thread $\tid$ has linearized (and is, hence, in \emph{done} state). 
}

From the point of view of a thread which \emph{does} the helping, the linearization condition forces its proof to scan over the pool of uncommitted triples registered in the helping protocol and identify those that need to be linearized at its linearization point\np{, changing their protocol state from \emph{pending} to \emph{done}. This step involves a proof obligation for the helping thread to show that the sequential specification of $\tid$'s operation is indeed satisfied at the linearization point.} 

\np{
One crucial innovation in our helping protocol is that we have formulated a linearization condition that is parametric in the sequential specification of the data structure operations, making the soundness proof for the hindsight specification applicable to many structures at once. In particular, we deal with the aspect of scanning and updating the registry in the proof of the helping thread,}
the proof author simply invokes a lemma provided by our method at the identified linearization points. Therefore, the helping protocol mechanism remains fully opaque to the proof author.




\subsection{Invariant for Hindsight Reasoning}
\label{sec-hindsight-invariant}

\newcommand{\Snap}{\mathbb{S}}
\newcommand{\snapshotcons}{\mathsf{snapshot\_constraints}}
\newcommand{\persnap}{\mathsf{per\_snapshot}}
\newcommand{\transinv}{\mathsf{transition\_inv}}
\newcommand{\resources}{\mathsf{resources}}
\newcommand{\pastt}[1]{\past_{#1}}

Hindsight arguments involve reasoning about past program states. Our encoding therefore tracks information about past states using \emph{computation histories}. 
We define computation histories as finite partial maps from \emph{timestamps}, $\Nat$, to \emph{snapshots}, $\Snap$. A snapshot describes an abstract view of a program state. It is a parameter of our method. For instance, a snapshot may capture the physical memory representation of the data structure under proof, while abstracting from the remainder of the program state.
Another parameter is a function $\cssabs{\cdot}$ that computes the abstract state of the data structure from a given snapshot. 

\begin{figure}[h]
  \begin{align*}
    \mainInv(r) \defeq {} & \exists\, H\, T\, C.\; \csslstate(r, C) * \cssabs{H(T)} = C\\
    {} & * \historyinv(H, T)  \,*\, \helpinginv(H, T) \,*\, \tplinv(r, H, T) \\
    \tplinv(r, H, T) \defeq {} & \resources(r, H(T))\\
    {} & * (\forall t,\; 0 \le t \le T \impl \persnap(H(t)))\\
    {} & * (\forall t,\; 0 \le t < T \impl \transinv(H(t), H(t+1)))
  \end{align*}
  \caption{Definition of the shared state invariant encoding the hindsight reasoning. Variable $H$ represents the history, $T$ the current timestamp in use and $C$ the abstract state of the structure.}
  \label{fig-helping-prot-inv}
\end{figure}

\refFig{fig-helping-prot-inv} shows a simplified definition of the invariant that encodes the hindsight reasoning. For sake of brevity, we provide only a high-level overview of the predicates used in the invariant. The predicate $\historyinv(H, T)$ contains the mechanism to track the history of snapshots. That is, $H$ denotes the history that has been observed so far and $T$ is the current time stamp. Using appropriate ghost resources, it ensures that the timestamps are non-decreasing and past states recorded in $H$ are preserved by future updates to the history. This allows us to define a \emph{past predicate} $\pastt{s, t_0} (q)$ with the intuitive meaning that the history contains state $s$ recorded after (or at) time $t_0$ for which proposition $q$ holds true. The exact definition of the past predicate uses the ghost resources used to preserve the past states. The predicate $\historyinv(H, T)$ also guarantees that $\dom(H) =  \{0 \dots T\}$, ensuring that there are no gaps in the history. 

The conjunct $\abs{H(T)} = C$ and the predicate $\csslstate(r,C)$ together tie the abstract state $C$ of the data structure to the latest snapshot in the history. The predicate $\csslstate(r,C)$ is the dual of the representation predicate $\cssstate(r,C)$ used in the client-level atomic specification. Both represent one half of an ownership over the abstract state of the structure, keeping the abstract state defined by $\mainInv(r)$ synchronized with the representation predicate $\cssstate(r,C)$. 

The helping protocol predicate $\helpinginv(M, T)$ contains a \emph{registry} of thread IDs with unmodifying linearization points that require helping from other concurrent threads. For each thread ID $\tid$ in the registry, the protocol stores information such as the start time of the thread, whether it has been linearized or not, etc.

The predicate $\tplinv(r, H, T)$ captures invariants particular to the data structure under proof. It is further composed of three abstract predicates that are meant to be instantiated with the structure specific invariants. The three predicates serve the following purpose. The first predicate $\resources(r, H(T))$ ties the current snapshot to the physical representation of the structure. The predicate $\historyinv(H, T)$ contains a conjunct $(\forall t, t < T \impl H(t) \neq H(t+1))$. Together with the predicate $\resources$, this conjunct forces a thread to update the history whenever the structure is modified. 

The predicate $\persnap(H(T))$ captures the structural invariants that hold for any given snapshot. For instance, when proving the skiplist template, this predicate holds facts about the nodes $\vhead$ and $\vtail$ having maximum height, etc. The predicate $\transinv(s, s')$ captures a transition invariant on snapshots observed in the history. That is, it constrains how certain quantities evolve over time. Again as an example from the skiplist template proof, the fact that a node marked in $s$ remains marked in $s'$ is included here. Crucially, the facts in $\transinv(s, s')$ allow temporal interpolation required to establish facts about past states in the history (like in \cref{sec-hindsight-intro}).

To summarize, the proof author defines the snapshot of the structure, the function $\cssabs{\cdot}$, and instantiates the three abstract predicates in $\tplinv$ appropriately. The resulting shared state invariant then tracks the history and handles the helping protocol without requiring further fine-tuning to the data structure at hand.


\section{Verifying the Skiplist Template}
\label{sec-template-proof}

We relate the intuitive proof argument from \cref{sec-proof-intuition} to the development on hindsight reasoning in Iris in \cref{sec-hindsight} to obtain a complete proof of the skiplist template.  To achieve this, we must perform three tasks required by the proof method in \cref{sec-hindsight}. The first task is to determine the decisive operations that potentially alter the structure, and resolve the prophecy around those operations. As discussed previously, the decisive operations are \code{markNode} for \code{delete} and \code{changeNext} for \code{insert}. The \code{search} operation does not modify the abstract state and hence, it has no decisive operation. 

The second task is to define a snapshot in the context of the skiplist template and instantiate $\tplinv$ appropriately. This includes the predicate $\resources$ that ties the concrete state of the structure to the latest snapshot, as well as invariants that allow temporal interpolation. The third and the final task is to prove the hindsight specification for the core operations.

In this section we focus on the second task of defining the snapshot and providing invariants necessary to formalize the intuitive proof argument. Once, we have set up the right invariants, the formalized proof follows the intuitive proof very closely. We explain this with \code{delete} as an example. 

\subsection{Snapshot and the Skiplist Template Invariant}

Recall that the notion of keysets are central to the intuitive proof argument for the core operations of the skiplist template. 
Hence, a snapshot of the structure must contain information about the keysets. For encoding keysets in Iris, we borrow heavily from \cite{DBLP:conf/pldi/KrishnaPSW20}, especially the \emph{keyset camera} and the representation of keysets via the Flow Framework. \more{We defer further details on this matter to \cref{sec-template-proof-extra}.}

We define the snapshot of the skiplist template as a tuple containing the following components:
\begin{itemize}
\item the set of nodes $N$ comprising the structure (also referred to as the \emph{footprint} below)
    \item the abstract state of the structure (a set of keys)
    \item the mark bits (a map from $N$ to $\Nat \to \mathsf{Bool}$, i.e., a Boolean per level)
    \item the next pointers (a map from $N$ to $\Nat \to N$)
    \item the keys (a map from $N$ to $\keyspace$)
    \item the height of nodes (a map from $N$ to $\Nat$)
    \item the representation of flow values 
\end{itemize}

\newcommand{\fpof}[1]{\m{FP}(#1)}

We reparameterize the $\markof{n, i}$ function introduced earlier to take the snapshot as an argument. Thus, we use $\markof{s, n, i}$ to mean the mark bit of node $n$ at level $i$ in snapshot $s$. We redefine $\nextof{\cdot}$, $\keyof{\cdot}$, $\keysetof{\cdot}$ and other such functions similarly by adding the snapshot $s$ as an additional parameter. We also use $\fpof{s}$ to represent the footprint of the snapshot $s$.

\newcommand{\vmark}{\mathit{mk}}
\newcommand{\vnext}{\mathit{nx}}
\newcommand{\vmarkp}{\mathit{mk}'}
\newcommand{\vnextp}{\mathit{nx}'}
\newcommand{\resourcesks}{\mathsf{resources\_keyset}}

\begin{figure}

  \begin{align*}
    \tplinv(r, H, T) \defeq {} & \resources(r, H(T))\\
    {} & * (\forall t,\; 0 \le t \le T \impl \persnap(H(t)))\\
    {} & * (\forall t,\; 0 \le t < T \impl \transinv(H(t), H(t+1)))\\
    \resources(s) \defeq {} & \Sep_{n \in \fpof{s}} \nodeInv(n, \markof{s, n}, \nextof{s, n}, \keyof{s, n}, \htof{s, n}) \\
    {} & * \resourcesks(s)\\
    \transinv(s, s') \defeq {} & (\fpof{s} \subseteq \fpof{s'}) \\
    {} & * (\forall n,\; \keyof{s', n} = \keyof{s, n} \land \htof{s', n} = \htof{s, n}) \\
    {} & * (\forall n\; i,\; \markof{s, n, i} = \true \impl \markof{s', n, i} = \true) \\
    {} & * (\forall n\; i,\; \markof{s, n, i} = \true \impl \nextof{s', n, i} = \nextof{s, n, i})
  \end{align*}
  \caption{Instantiating $\tplinv$ with invariants of the skiplist template.}
  \label{fig-template-tplinv}
\end{figure}

We now present the skiplist template invariant in \cref{fig-template-tplinv}. The $\resources$ predicate ties the snapshot to the concrete state through an intermediary node-level predicate $\nodeInv(n, k, h, \vmark, \vnext)$. This predicate actually ties the physical representation of a node in the heap to the abstract quantities ($\keyof{\cdot}$, $\htof{\cdot}$, $\markof{\cdot}$ and $\nextof{\cdot}$, respectively) that the skiplist template relies on. The $\nodeInv$ predicate also owns all the resources needed to execute the helper functions. The skiplist template proof is parametric in the definition of $\nodeInv$. Thus, we achieve proof reuse across skiplist variants that follow the same high-level skiplist algorithm, but implement the node differently. We provide more details on this matter later. We discuss some concrete node implementations in \cref{sec-evaluation}.

The predicate $\resourcesks(s)$ capture the ownership resources required for keyset reasoning. Using the ghost resources in Iris and the keyset camera from \cite{DBLP:conf/pldi/KrishnaPSW20}, it ensures that the keysets and the logical contents of nodes in $s$ satisfy $(\ref{eqn-keyset})$. 


The predicate $\persnap$ captures structural invariants that hold for all snapshots recorded in the history. This includes invariants of three kinds: first, invariants to ensure that each component of the snapshot is of the correct type and the maps (from nodes to mark bits, next pointers, etc.) are defined for all nodes in the footprint; second, the node-level invariants relating the node's inset, outset, mark bit, etc (like Invariant~\ref{inv-sk1}); and third, invariants about the $\vhead$ and $\vtail$ nodes, such as $\keyof{s, \vhead} = -\infty$, $\htof{\vtail} = L$, etc. 

The predicate $\transinv(s, s')$ captures invariants about how certain quantities evolve over time, such as that  mark bits once set to true remain true. The invariants~\ref{inv-sk2}, \ref{inv-sk3}, and \ref{inv-sk6} presented in \cref{sec-proof-intuition} are part of this predicate. These invariants form the crux of the hindsight reasoning, as they enable temporal interpolation.

Before we go into the formal proof argument for \code{delete}, we must discuss how to reason about the node-level helper functions. \Cref{fig-template-helper-specs} shows the specification for the helper functions assumed by the skiplist template. The specifications are logically atomic, i.e., they behave like a single atomic step in the template. The preconditions for all of the functions rely solely on the predicate $\nodeInv$. The functions \code{getKey}, \code{getHeight} and \code{findNext} read various components of the node. Note that \code{findNext} reads both the mark bit and the next pointer together.

The specification for functions \code{markNode} and \code{changeNext} is slightly more complex because they potentially change the structure. Let us explain them briefly. For \code{markNode} on node $n$ at level $i$, the return value (\code{Success} or \code{Failure}) is determined by whether $n$ is already marked at $i$. If it is, then the function returns \code{Failure} without modifying the node. If it is unmarked, then \code{markNode} successfully marks it, and updates the node accordingly. The specification for \code{changeNext} can be interpreted similarly. Here, the return value hinges upon the mark bit being false and the next pointer of $n$ pointing to $n'$ at $i$.

\begin{figure}
  \centering
  \begin{lstlisting}
$\annotAtom{k\,h\,\vmark\,\vnext.\; \nodeInv(n, k, h, \vmark, \vnext)}$ |<getKey>| $n$ $\annotAtom{\Ret k.\; \nodeInv(n, k, h, \vmark, \vnext)}$
$\annotAtom{k\,h\,\vmark\,\vnext.\; \nodeInv(n, k, h, \vmark, \vnext)}$ |<getHeight>| $n$ $\annotAtom{\Ret h.\; \nodeInv(n, k, h, \vmark, \vnext)}$
$\annotAtom{k\,h\,\vmark\,\vnext.\; \nodeInv(n, k, h, \vmark, \vnext) * (i < h)}$ |<findNext>| $i$ $n$ $\annotAtom{\Ret n'.\; \nodeInv(n, k, h, \vmark, \vnext) * (\vnext(i) = n')}$

$\annotAtom{k\,h\,\vmark\,\vnext.\; \nodeInv(n, k, h, \vmark, \vnext) * (i < h)}$ |<markNode>| $i$ $n$
$\annotAtom{\Ret x.\; \nodeInv(n, k, h, \vmarkp, \vnext) * (\vmark(i) = \true \impl x = \mathsf{Failure} * \vmarkp = \vmark) \\ * (\vmark(i) = \false \impl x = \mathsf{Success} * \vmarkp = \vmark[i \rightarrowtail \true]) }$

$\annotAtom{k\,h\,\vmark\,\vnext.\; \nodeInv(n, k, h, \vmark, \vnext) * (i < h)}$ |<changeNext>| $i$ $n$ $n'$ $e$
$\annotAtom{\Ret x.\; \nodeInv(n, k, h, \vmark, \vnextp) * ((\vmark(i) = \true \lor \vnext(i) \neq n') \impl x = \mathsf{Failure} * \vnextp = \vnext) \\ * ((\vmark(i) = \false \land \vnext(i) = n') \impl x = \mathsf{Success} * \vnextp = \vnext[i \rightarrowtail e])}$
\end{lstlisting}
\caption{Specifications of the helper functions used by the skiplist template.}
\label{fig-template-helper-specs}
\end{figure}

\subsection{Proof of \code{delete}}

We now have all the ingredients to show that \code{delete} satisfies $(\ref{eq:hindsight-spec})$. \tw{We  provide only a high-level summary of the proof here. Please see \cref{sec-template-proof-extra-delete} for more details.}

The precondition provides access to the invariant $\mainInv(r)$ and knowledge that the thread ID is $\vtid$ with start time $t_0$. Additionally, the thread has the right to resolve prophecy $p$ around the decisive operations, and if the thread observes a successful decisive operation, then the atomic update $\atomicUpdate(\Phi)$ is available to help with the linearization. The \code{delete} operation begins with \code{traverse}. Using the $\past$ operator defined in \cref{sec-hindsight-invariant}, we express the postcondition of \code{traverse} as 
\[ \pastt{s, t_0}(k \in \keysetof{s, \vcurr} \land (\vres \iff k \in \contentsof{s, \vcurr})).\]
Intuitively, this assertion captures that there is a past state $s$ in the history (after time point $t_0$) in which $k$ is in the keyset of $\vcurr$ and $\vres$ is true iff $k$ is in the logical contents of $\vcurr$. 

The argument here, like the intuitive proof argument in \cref{sec-proof-intuition} proceeds by case analysis on $\vres$ (returned by \code{traverse}). Let us first consider the case that $\vres$ is $\false$. The \code{delete} operation also terminates with $\false$. Since the thread terminates without any calls to the decisive operations, this case corresponds to the $\neg \procProph(\vpvs)$ case in the postcondition of (\ref{eq:hindsight-spec}). The postcondition requires \code{delete} to establish the predicate $\pastlin(\code{del}, k, \false, t_0)$. In this context, establishing this predicate amounts to identifying a witness past state in which $k$ was not part of the abstract state. Clearly, this is witnessed by state $s$ from the specification of \code{traverse}. Applying (\ref{eqn-keyset}) in state $s$, we can establish the predicate $\pastlin(\code{del}, k, \false, t_0)$.

Now, let us consider the case that $\vres$ is $\true$. The \code{maintainanceOp\_del} marks node $\vcurr$ at the higher level, but the interesting part of the proof is when the decisive operation \code{markNode} is called at the base level (\cref{line-delete-markNode}). Again there are two cases to consider, depending on whether \code{markNode} succeeds. If \code{markNode} succeeds, then we can establish $\procProph(\vpvs)$ as we see a \code{Success} value being resolved. In this case, the precondition of $(\ref{eq:hindsight-spec})$ provides the atomic update $\atomicUpdate(\Phi)$. Since, the thread has modified the abstract state, this becomes the linearization point. The thread can linearize with $\atomicUpdate(\Phi)$ to obtain the receipt $\Phi$ and satisfy its postcondition. The proof also has to update the history with the new snapshot of the structure, as $\vcurr$ goes from being unmarked to marked.

The final (and most interesting) case is when \code{markNode} fails. Here again, we must establish $\pastlin(\code{del}, k, \false, t_0)$ to complete the proof of $(\ref{eq:hindsight-spec})$. Two facts are useful: (i) in the past state $s$ referred to in the \code{traverse} spec, we can establish that $\markof{s, \vcurr} = \false$; and (ii) since the \code{markNode} has failed, in the current state say $s_0$, $\markof{s_0, \vcurr} = \true$. Hence, by using the second conjunct of $\transinv$ in \cref{fig-template-tplinv} and temporal interpolation on the two facts above, we can infer the existence of two consecutive states $s_1$ and $s_2$, such that $\markof{s_1, \vcurr} = \false$ and $\markof{s_2, \vcurr} = \true$. Clearly, a concurrent \code{delete} thread marked $\vcurr$ in state $s_2$. Hence, this state becomes the witness to establish the predicate $\pastlin(\code{del}, k, \false, t_0)$.  This completes the proof that \code{delete} satisfies $(\ref{eq:hindsight-spec})$.


\section{Proof Mechanization and Evaluation}
\label{sec-evaluation}


We now shed light on the mechanization of the hindsight methodology, as well as its application to the skiplist template. We additionally reverify the multicopy template from~\cite{DBLP:journals/pacmpl/PatelKSW21} using our new hindsight specification to modularize the proof effort. 
\npout{This case study allows us to (i) showcase the general applicability of our hindsight method in Iris and (ii) compare the proof effort of hindsight-based arguments against tailor-made arguments that are specific to the data structure at hand.}
\np{Although the multicopy algorithms are lock-based, hindsight reasoning is helpful in their verification. The case study demonstrates a substantial reduction in proof size due to the encoding of hindsight reasoning in Iris, illustrating the generality of our contribution.}
\npout{We plan to submit our development for artifact evaluation.}
\np{Our development is available as a VM and docker image on Zenodo~\cite{lockfree-templates-artifact}.}

All of the proofs we discuss below are mechanized in Iris/Coq. The templates, traversals and the node implementations are written in Iris's default programming language HeapLang. In order to correctly capture the dependence between different layers of the proofs (such as hindsight specification and the templates, the templates and the \code{traverse}/node implementations), we heavily make use of Coq's module system.

\begin{figure}[t]
    \centering
    \begin{tikzpicture}[>=stealth, scale=0.8, every node/.style={scale=0.75}]
      \tikzstyle{gnode}=[rectangle, rounded corners=2pt, draw=black, fill={rgb,255:red,198;green,222; blue,240}, thick, minimum height=0.7cm, minimum width=3.2cm, font=\normalsize]
  
      \def\xsep{1.2}
      \def\ysep{-1.5}
      \def\xshift{5}
      \def\yshift{-2.5}
  
      \node (imp) {\begin{tabular}{ c c } 
        \scalebox{1.5}{$\longrightarrow$} & \scalebox{1.5}{$\dashrightarrow$} \\
        \large{satisfies} & \large{assumes} \\
        \end{tabular}};
      \node[gnode, fill=white] (hs) at ($(imp) + (0, \ysep)$) {Hindsight};
      \node[gnode, fill=white] (cs) at ($(hs) + (0, \ysep)$) {Client-level Spec};
  
      \node[gnode] (nd) at ($(hs) + (\xshift, -\ysep)$) {Node};
      \node[gnode] (tr) at ($(nd) + (0, \ysep)$) {Traverse};
      \node[gnode] (sk) at ($(tr) + (0, \ysep)$) {Skiplist Template};
      \node[gnode, fill={rgb,255:red,210;green,231;blue,206}] (mc) at ($(sk) + (0, \ysep)$) {Multicopy Template};
  
      \node[gnode] (nd1) at ($(nd) + (\xshift, 0)$) {Node Impl. 1};
      \node[gnode] (nd2) at ($(nd1) + (0, \ysep)$) {Node Impl. 2};
      \node[gnode] (eg) at ($(nd2) + (0, \ysep)$) {Eager Travresal};
      \node[gnode] (lz) at ($(eg) + (0, \ysep)$) {Lazy Travresal};
  
      \draw[edge, dashed] (hs) to (cs);
      \draw[edge] (sk.west) to (hs.south east);
      \draw[edge] (mc.north west) to (hs.south east);
      \draw[edge, dashed] (nd) to (tr);
      \draw[edge, dashed] (tr) to (sk);
      \draw[edge] (nd1.west) to (nd.east);
      \draw[edge] (nd2.north west) to (nd.east);
      \draw[edge] (eg.north west) to (tr.south east);
      \draw[edge] (lz.north west) to (tr.south east);
  
    \end{tikzpicture}
    \caption{The structure of our proofs. Each box represents a collection of modules relevant to the label. The dashed arrows represent module dependence, i.e., assumption of specifications. The normal arrows represent implementation of the target module (fulfillment of the assumptions).}
    \label{fig-proof-organization}
  \end{figure}
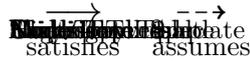

The organization of our proofs is shown in \Cref{fig-proof-organization}. Going from left to right, the first column relates to the formalization of hindsight reasoning in Iris. The box ``Hindsight'' captures the assumptions regarding the hindsight specification from \Cref{sec-hindsight}. These assumptions not only include the hindsight specification itself but also the relevant definitions of snapshots, histories, etc. The module ``Client-level Spec'' relates the client-level specification expressed in terms of atomic triples to the hindsight specification used for the template-level proofs. The corresponding proof involves the reasoning about prophecies and the helping protocol, which is done once and for all and applicable to all data structures that fulfill the assumptions made in the ``Hindsight'' module.


The middle column consists of modules for the two verified templates (skiplist and multicopy) and the associated proofs verifying the template operations against the hindsight specification. We discuss them in turn.

\smartparagraph{Skiplist template case study}
The skiplist template, as described in \Cref{fig:skiplist-template}, abstracts from the concrete implementations of nodes and the \code{traverse} operation. Hence, we package their specifications into separate modules. To ensure that the specified data structure invariant for the skiplist template is not vacuous, we also verified an \code{init} routine that initializes the data structure and establishes the invariant.

The final column shows modules for the two node implementations of the skiplist template, as well as the eager and lazy traversal discussed in \Cref{sec-templates}. The helper functions \code{markNode} and \code{changeNext} are implemented using an atomic CAS operation in both of the node implementations. The crux of the node implementation for the skiplist template is to determine a memory representation of the mark bit and the next pointer (at some level) such that both values can be read or written together with one atomic CAS operation. The first node implementation does this by using a sum type
. The second node implementation is conceptually similar but uses more low-level data types instead of a sum type.

\np{The traversal and node implementations above correspond to several existing lock-free (skip)list algorithms from the literature. The Herlihy-Shavit skiplist algorithm~\cite[\S~14]{DBLP:books/daglib/0020056} is obtained by instantiating our template with the eager traversal, the node implementation 2, and maintenance operations that link higher-level nodes in increasing order of level and  unlink nodes in the opposite order. The Michael set~\cite{DBLP:conf/spaa/Michael02} is obtained as a degenerate case of the Herlihy-Shavit template instantiation where the skiplist is restricted to $\vL=2$ (For $\vL=2$, Level $1$  consists of only a fixed single edge between the sentinel nodes. So, conceptually, Level 1 can be ignored in this case.)

We obtain a novel variant of a skiplist by replacing the eager traversal in the Herlihy-Shavit instantiation with the lazy traversal. The lazy traversal is inspired by the Harris list algorithm~\cite{DBLP:conf/wdag/Harris01}, which is obtained as a degenerate case of this new lazy skiplist algorithm by restricting it to $\vL=2$.}

We present a summary of the proof effort for the skiplist template in \cref{tab-skiplist-case-study}. The proof-checking time was measured on the Docker image running on an Apple M1 Pro chip with 16GB RAM. The flow library contains the Iris formalization of the Flow Framework developed in~\cite{DBLP:conf/pldi/KrishnaPSW20, DBLP:journals/pacmpl/PatelKSW21}. As a minor contribution, we extend this library with general lemmas for reasoning about graph updates that have an affect on an unbounded number of nodes. These lemmas are useful for the proofs of \code{insert}, \code{delete} and lazy \code{traverse}. The unbounded updates, as well as the maintenance operations, are the reason for the relatively high number of proof lines for the \code{insert} and \code{delete} operations.

\begin{table}[t]
  \centering
  \setlength{\tabcolsep}{5pt}

  \begin{tabular}[t]{l r r r r}
    \multicolumn{5}{l}{\bf Skiplist Template (Iris/Coq)}\\[2pt]
    \hline
    \bf Module & \bf Code & \bf Proof & \bf Total & \bf Time \\
    \hline
      Flow Library    & 0     & {\the\numexpr 5330-0}     & 5330  & 33\\
      Hindsight   & 0  & {\the\numexpr 950-0}     & 950   & 11\\
      Client-level Spec   & 9     & {\the\numexpr 338-9}     & 338   & 18\\
      Skiplist             & 12       & {\the\numexpr 1705-12}    & 1705    & 26\\
      Skiplist Init($*$)         & 6       & {\the\numexpr 325-6}    & 325    & 15\\
      Skiplist Search($*$)       & 7     & {\the\numexpr 69-7}    & 69   & 6\\
      Skiplist Insert($*$)         & 37     & {\the\numexpr 3494-37}    & 3494   & 111\\
      Skiplist Delete($*$)        & 28     & {\the\numexpr 2429-28}  & 2429  & 72\\
      Node Impl. 1           & 118   & {\the\numexpr 1026-118}   & 1026   & 35\\
      Node Impl. 2      & 106   & {\the\numexpr 942-106}   & 942   & 35\\
      Eager Traversal   & 38      &  {\the\numexpr 1203-38}  &  1203 & 96\\
      Lazy Traversal   & 47      & {\the\numexpr 2110-47}   &    2110& 145\\
      \bf Total               & \bf {\the\numexpr 0+0+9+12+6+7+37+28+118+106+38+47}   & \bf {\the\numexpr 19921-408}   & \bf {\the\numexpr 5330+950+338+1705+325+69+3494+2429+1026+942+1203+2110}   & \bf {\the\numexpr 33+11+18+26+15+6+111+72+35+35+96+145}\\
    \hline
    \hline
      Herlihy-Shavit & {\the\numexpr 0+9+12+6+7+37+28+106+38} & {\the\numexpr 11455-243} & {\the\numexpr 950+338+1705+325+69+3494+2429+942+1203} & {\the\numexpr 11+18+26+15+6+111+72+35+96} \\
    \hline
  \end{tabular}
  \caption[Summary of proof effort]{
  Summary of the proof effort. For each module, we show the number of lines of program code, lines of proof, total number of lines, and the proof-checking time in seconds. The code for the initialization and the core operations of the skiplist (entries with ($*$)) is technically defined in the ``Skiplist'' module, however here we present them separately for each operation to provide a better picture. The count for Herlihy-Shavit is the summation of rows ``Hindsight'', ``Client-level Spec'', all ``Skiplist'' modules, ``Node Impl. 2'' and ``Eager Traversal''.}
  \label{tab-skiplist-case-study}
\end{table}

\smartparagraph{Multicopy template case study}
The multicopy template from~\cite{DBLP:journals/pacmpl/PatelKSW21} generalizes search structures such as the lock-based Log-Structured Merge (LSM) tree used widely in modern database systems. It satisfies the Map ADT specification, with \code{search} and \code{upsert} (for insert/update) as its core operations. To deal with the complexity of future-dependent external linearization points, the original proof relies on an intermediate template-level specification based on the concept of \emph{search recency}.


\np{\Cref{tab-multicopy-case-study} presents a detailed comparison of the multicopy template proofs from~\cite{DBLP:journals/pacmpl/PatelKSW21} versus the new proof based on the hindsight framework. The original proof consists of a total of 2779 lines. By contrast, the definitions (``Defs'') and ``Client-level Spec'' proofs can be factored out of the total cost of the hindsight-based proof, because it is part of the hindsight library itself. Hence, the new proof based on hindsight reasoning  consists of only {\the\numexpr 540+399+371} lines, which is a reduction of 53\%. To summarize, the improvement stems from the fact that the original proof relies on an intermediate specification and a helping protocol that is tailored to multicopy structures, while our new proof uses a helping protocol that is shared among all proofs that build on the new hindsight proof method.

While the majority of the reduction in the proof size stems from the elimination of structure-specific specifications and helping protocol proofs, we also saw a minor reduction in the size of the remainder of the proof. One outlier is the proof of \code{upsert}. Here, the increase is attributed to the fact that the proof has to construct a fresh snapshot when the operation succeeds. However, this construction is conceptually simple and could be factored out into more abstract lemmas that are provided directly by the hindsight library. }

\begin{table}[t]
  \centering
  \setlength{\tabcolsep}{5pt}

  \begin{tabular}[t]{l r r}
    \multicolumn{3}{l}{\bf Multicopy Template (Iris/Coq)}\\[2pt]
    \hline
    \bf Module & \bf Original & \bf Hindsight \\
    \hline
      Defs & 866 & (950) \\
      Client-level Spec & 434 & (338) \\
      LSM & 741 & 540 \\
      Search & 411 & 399 \\
      Upsert & 327 & 371 \\
      \bf Total & \bf {\the\numexpr 866+434+741+411+327} & \bf {\the\numexpr 540+399+371} \\
    \hline
  \end{tabular}
  \caption{Comparison of multicopy template proofs. The column ``Original'' shows the number of lines from the proofs in~\cite{DBLP:journals/pacmpl/PatelKSW21}, while ``Hindsight'' shows them for our new proof effort. Module ``Defs'' represents definitions required for proving the client-level specification (helping invariant, history predicate, etc). Module ``Client-level Spec'' contains the proof relating the intermediate specification (Search Recency Specification from~\cite{DBLP:journals/pacmpl/PatelKSW21} and Hindsight Specification in our paper) to the client specification. Module ``LSM'' contains definitions required to instantiate the frameworks for LSM trees. Modules ``Search'' and ``Upsert'' refer to the proofs for the search and upsert operations, respectively. Entries in `()' for the `Hindsight' column are not included in the total due to being part of the hindsight library. }
  \label{tab-multicopy-case-study}
\end{table}




\section{Related Work}



The formal verification of linearizability has received much attention in recent years.
We refer to~\cite{DBLP:journals/csur/DongolD15} for a survey of relevant techniques and focus our discussion to the most closely related works.

Our work builds on the idea of template algorithms for lock-based concurrent search structures of~\cite{DBLP:series/synthesis/2021Krishna,DBLP:journals/pacmpl/PatelKSW21,DBLP:conf/pldi/KrishnaPSW20}, which we extend to the setting of lock-free implementations. A common challenge when verifying linearizability of lock-free data structures is the prevalence of future-dependent and external linearization points. Hindsight theory~\cite{DBLP:conf/podc/OHearnRVYY10,DBLP:conf/wdag/Lev-AriCK15,DBLP:conf/wdag/FeldmanE0RS18,DBLP:journals/pacmpl/FeldmanKE0NRS20, DBLP:journals/pacmpl/MeyerWW22,DBLP:journals/pacmpl/0001W023} has emerged as a suitable technique to address this challenge in the context of concurrent search structures. To our knowledge, we are the first to formalize hindsight reasoning within a foundational program logic. Tools like \tool{Poling}~\cite{DBLP:conf/cav/ZhuPJ15}, \tool{plankton}~\cite{DBLP:journals/pacmpl/MeyerWW22,DBLP:journals/pacmpl/0001W023}, and \tool{nekton}~\cite{DBLP:conf/cav/MeyerOWW23} automate hindsight reasoning at the expense of an increased trusted code base. However, these tools currently cannot handle complex data structures with unbounded outdegree like skiplists. Also, they do not aim to characterize the design space of related concurrent data structures like our template algorithms do.

Other techniques for dealing with future-dependent linearization points include arguments based on forward simulation (e.g., by tracking all possible linearizations of ongoing operations~\cite{10.1145/3632924}, tracking a partial order~\cite{DBLP:conf/esop/KhyzhaDGP17}, or using commit points~\cite{DBLP:conf/cav/BouajjaniEEM17}) and backward simulation (e.g., using prophecy variables~\cite{DBLP:journals/tcs/AbadiL91,DBLP:conf/pldi/LiangF13,DBLP:journals/pacmpl/JungLPRTDJ20}). Our encoding of hindsight reasoning in Iris combines forward reasoning (by tracking the history of the data structure state) and backward reasoning (by using prophecies). However, the details of this encoding are for the most part hidden from the proof engineer by providing a higher-level reasoning interface based on past predicates and temporal interpolation as proposed in~\cite{DBLP:journals/pacmpl/0001W023}. Our comparison with a prior proof of multicopy structure templates~\cite{DBLP:journals/pacmpl/PatelKSW21} suggests that this abstraction helps to reduce the proof complexity.

Several works propose techniques for automatically verifying concurrent skiplists. Abdulla et al.~\cite{DBLP:conf/atva/AbdullaHJLTV13} propose a technique for verifying linearizability of lock-free list-based data structures using forest automata. The evaluation considers bounded skiplists with up to 3 levels. However, the implementation does not scale to larger bounds and the unbounded case is outside the scope of the technique. 
\np{We note that the height of the skiplist is tied to the expected runtime of the skiplist operations. To guarantee the expected worst-case runtime bounds, the skiplist's height must be of order $O(\log(n))$ where $n$ is the expected maximal number of entries in the list. For this reason, real-world skiplist implementations are also parametric in the height. Heights up to 63 levels are feasible in deployed skiplists~\cite{folly-concurrent-skiplist}, so the restriction to height 3 in~\cite{DBLP:conf/atva/AbdullaHJLTV13} is unrealistic. By contrast, our proofs cover skiplists of arbitrary height.}

S\'anchez and S\'anchez~\cite{DBLP:conf/atva/SanchezS14} present an SMT-based approach towards an automated verification of concurrent lock-based skiplists. The approach is based on a decidable theory of unbounded skiplists. However, it does not consider lock-free implementations and focuses on establishing \emph{shape invariants} preserved by the structure instead of proving linearizability.

Unlike these automated tools, our approach does not rely on data-structure specific decidable theories for reasoning about inductive properties of heap graphs. Instead, we build on the Flow Framework~\cite{DBLP:journals/pacmpl/KrishnaSW18,DBLP:conf/esop/KrishnaSW20,DBLP:conf/tacas/MeyerWW23}, which enables local reasoning about such properties over general graphs in separation logic. As a minor contribution, we extend the mechanization of the Flow Framework from~\cite{DBLP:series/synthesis/2021Krishna} with lemmas to reason about graph updates that affect properties of an unbounded number of nodes.

\np{There are some skiplist algorithms that are not immediately covered by our template algorithm. For example, skiplists based on the algorithm presented in~\cite{DBLP:conf/podc/FomitchevR04} such as Java's \code{ConcurrentSkipListMap}~\cite{java-skiplist-set} use \emph{backlinks} to avoid restarts when a traversal fails. However, we believe that our template algorithm can be extended to subsume such algorithms by abstracting from the restart policy, similarly to how the present template abstracts from the maintenance policy.}

In this paper, we assume a programming language with a garbage collected semantics. The rationale for this assumption is that issues arising from manual memory reclamation can be addressed by orthogonal means. For instance, \cite{DBLP:journals/pacmpl/MeyerW19, DBLP:journals/pacmpl/MeyerW20} propose a technique that decouples the proof of data structure correctness from that of the underlying memory reclamation algorithm, allowing the correctness proof of the data structure to be carried out under the assumption of garbage collection. Recent work also showed how to carry out such modular proofs in program logics like Iris~\cite{DBLP:journals/pacmpl/JungLCKPK23}.


\section{Conclusions and Future Work}

This paper shows how to verify some of the most challenging concurrent data structure algorithms in existence. The accompanying proofs are fully mechanized in the foundational program logic Iris. The proofs are modular and cover the broader design space of the underlying algorithms by parameterizing the verification over aspects such as the low-level representation of nodes and the style of data structure maintenance.

Besides being the first work to verify unbounded lock-free skiplists, the work has developed technologies for Iris, particularly hindsight reasoning, that can be useful in many applications.

Our proofs guarantee safety but not liveness. This limitation is shared by the algorithms they verify: in any highly concurrent (minimal or no locking) setting, a thread $t$ may never complete  because of other threads that overtake it. Fortunately, this never happens in practice where threads all advance more or less at the same pace. Verifying liveness under such fairness assumptions remains an interesting direction for future work.


Another area of future work is to verify algorithms that mix locking parts with lock-free parts both for single copy and multicopy search structures. We believe that the present  framework will be a good basis for that effort.

\paragraph*{Acknowledgments}
This work is funded in parts by NYU Wireless and by the United States National Science Foundation under
grants CCF-2304758, 1840761, 2304758, and 25-74100-F1202.   Further funding came from an Amazon Research Award Fall 2021. Any opinions, findings, and conclusions or recommendations expressed in this material are those of 
the authors and do not reflect the views of Amazon. We thank Sebastian Wolff for many insightful discussions and his suggestions to improve the presentation of the paper.


\bibliography{references,dblp}

\clearpage

\more{
\appendix

\section{Proof of Traversals}
\label{sec-traverse-proof}

We provide additional details on the proof of the eager traversal, as well as the implementation and proof of the lazy traversal.

\subsection{Proof of Eager Traversal}
\label{sec-eager-proof}

\newcommand{\eagerinv}{\mathsf{EagerInv}}

We describe the proof of $\eagerI$ in more detail. Let us formally define the traversal invariant $\eagerinv$ for $\eagerI$ as follows:
\[\eagerinv(k, \vpred, \vcurr) \defeq \past (\markof{\vpred, 0} = \false \land \keyof{\vpred} < k) * \past (k \in \insetof{\vcurr})\]
The predicate $\eagerinv$ captures the fact that $\eagerI$ witnessed node $\vpred$ to be unmarked and its key less than $k$ at some point during its traversal. Additionally, it witnessed that $k$ was in the inset  of $\vcurr$ at some prior state of the structure. We put the two facts about $\vpred$ and $\vcurr$ in separate $\past$ predicates because the two facts may hold for different past states.  

\begin{figure}[h]
  \centering
  \begin{lstlisting}[aboveskip=1pt,belowskip=0pt, frame=none]
$\annot{\eagerinv(k, \vpred, \vcurr)}$
let $\eagerI$ $0$ $k$ $\vpred$ $\vcurr$ =   
  match |<findNext>| $0$ $\vcurr$ with
  | $\vcurrnext$, $\true$ -> 
    match |<changeNext>| $0$ $\vpred$ $\vcurr$ $\vcurrnext$ with
    | Success -> @\label{eageri-proof-changeNext-success}@
      $\annot{\eagerinv(k, \vpred, \vcurr) * \markof{\vpred, 0} = \false * \nextof{\vpred, 0} = \vcurrnext * \keyof{\vpred} < k}$ 
      $\annot{\markof{\vpred, 0} = \false * \keyof{\vpred} < k * k \in \insetof{\vpred} * k \in \outsetof{\vpred} * k \in \insetof{\vcurrnext}}$
      $\annot{\eagerinv(k, \vpred, \vcurrnext)}$
      $\eagerI$ $0$ $k$ $\vpred$ $\vcurrnext$@\label{eageri-proof-rec-1}@
    | Failure -> traverse $\vparr$ $\vsarr$ $k$
  | $\vcurrnext$, $\false$ ->@\label{eager-proof-findNext-unmarked}@
    $\annot{\eagerinv(k, \vpred, \vcurr) * \markof{\vcurr, 0} = \false * \nextof{\vcurr, 0} = \vcurrnext}$
    $\annot{\eagerinv(k, \vpred, \vcurr) * \markof{\vcurr, 0} = \false * \nextof{\vcurr, 0} = \vcurrnext \\ 
    * (\keyof{\vcurr} < k \implies k \in \eagerinv(k, \vcurr, \vcurrnext)) \\ * (k \le \keyof{\vcurr} \implies \past (k \in \keysetof{\vcurr})) }$
    let $\vcurrk$ = |<getKey>| $\vcurr$ in
    if $kc$ < $k$ then
      $\annot{\eagerinv(k, \vcurr, \vcurrnext)}$
      $\eagerI$ $0$ $k$ $\vcurr$ $\vcurrnext$@\label{eageri-proof-rec-2}@
    else
      $\annot{\past (k \in \keysetof{\vcurr})}$
      let $\vres$ = ($kc$ = $k$ ? $\true$ : $\false$) in
      ($\vpred$, $\vcurr$, $\vres$)@\label{eager-tri-return}@
      $\annot{\past (k \in \keysetof{\vcurr} \land \vres \iff k \in \contentsof{\vcurr})}$
  \end{lstlisting}

\caption{Outline for the proof of the $\eagerI$ for the base level.}
\label{fig-eageri-proof}
\end{figure}

For the moment, assume that predicate $\eagerinv(k, \vpred, \vcurr)$ holds at the beginning of the function call $\eagerI\;0\;k\;\vpred\;\vcurr$. The proof outline in \cref{fig-eageri-proof} shows that this is sufficient to prove the \code{traverse} specification. We explain the critical steps of the proof in the following. 

The first critical step occurs when $\vcurr$ is found to be marked, and the subsequent \code{changeNext} call on $\vpred$ succeeds (Line~\ref{eageri-proof-changeNext-success}). Immediately after this step, we can establish that $\markof{\vpred, 0} = \false$ and $\nextof{\vpred, 0} = \vcurrnext$ due to the success of \code{changeNext}. The Invariant~\ref{inv-sk3} allows us to infer $\keyof{\vpred} < k$ from $\eagerinv(k, \vpred, \vcurr)$. In addition, we use the following structural invariant:
\begin{enumerate}[resume*]
  \item \label{inv-sk4} For all nodes $n$, if $\markof{n, 0}$ is set to $\false$ then $[\keyof{n}, \infty) \subseteq \insetof{n}$. As a consequence of the definition of edgesets and $\outsetof{n}$ (from \Cref{sec-edgeset-framework}), we additionally obtain $\outsetof{n} = (\keyof{n}, \infty)$. 
\end{enumerate}
Invariant~\ref{inv-sk4} helps us establish that $k \in \insetof{\vpred}$, $k \in \outsetof{\vpred}$ and since $\nextof{\vpred, 0} = \vcurrnext$, also $k \in \insetof{\vcurrnext}$. Using the rule that $\vprop$ implies $\past \vprop$ for an arbitrary proposition $\vprop$, we can establish $\eagerinv(k, \vpred, \vcurrnext)$ before the recursive call on Line~\ref{eageri-proof-rec-1}.

The second critical step is when \code{findNext} determines $\vcurr$ to be unmarked (Line~\ref{eager-proof-findNext-unmarked}). The proof goes by case analysis, depending on whether $\keyof{\vcurr} < k$ holds. First consider that this condition is true. Similar to the first critical step, we are able to establish $\eagerinv(k, \vcurr, \vcurrnext)$ here as we have all the relevant facts at this moment. The predicate $\eagerinv(k, \vcurr, \vcurrnext)$ is used for the recursive call on Line~\ref{eageri-proof-rec-2}. Now consider the case that $k \le \keyof{\vcurr}$. Here, we use $\past (k \in \insetof{\vcurr})$ provided by $\eagerinv(k, \vpred, \vcurr)$. Let $t$ be the time point when $k \in \insetof{\vcurr}$ held true. Since $\vcurr$ is unmarked currently, $\vcurr$ must also have been unmarked at time $t$ due to Invariant~\ref{inv-sk2}. And by Invariant~\ref{inv-sk4}, we can establish that $k \notin \outsetof{\vcurr}$ at time $t$. Thus, we conclude that also $k \in \keysetof{\vcurr}$ held at time $t$. Similarly we conclude that $\contentsof{\vcurr} = \{ \keyof{\vcurr} \}$ as $\vcurr$ is unmarked. Putting this all together, we obtain: 
\[k \in \contentsof{\vcurr} \iff k = \keyof{\vcurr} \iff \vres.\] 
This concludes the proof that predicate $\eagerinv$ is indeed an invariant for $\eagerI$ and is sufficient to prove the \code{traverse} specification.

In the previous proof, we assumed $\eagerinv$ holds when $\eagerI$ begins executing at the base level. The final piece remaining from the complete proof of the \code{traverse} specification is to show that $\eagerI$ collects enough information at higher levels to establish $\eagerinv$ before traversing the base level. We explain this below.

It is easy to see that for all higher levels $i$, $\eagerI$ can establish $\past (\markof{\vpred, i} = \false \land \keyof{\vpred} < k)$ when it returns $(\vpred, \_, \_)$. (In fact, $\eagerinv$ contains the exact same facts about $\vpred$ and the same argument as in the proof outline for the base level is applicable.) Thus, when a predecessor $\vpred$ is chose from the higher level to initiate $\eagerI$ at the base level, we know that $\past (\markof{\vpred, i} = \false \land \keyof{\vpred} < k)$ holds. The current node $\vcurr$ is chosen by calling $\code{findNext}\;0\;\vpred$ at Line~\ref{eager-rec-findNext} in \cref{fig:eager-traversal}. If \code{findNext} determines $\vpred$ to be unmarked, then we can establish $k \in \insetof{\vcurr}$ using the fact $\past (\keyof{\vpred} < k)$ and Invariant~\ref{inv-sk4}. However, if $\vpred$ is found to be marked, then it is a bit tricky to establish that $\past (k \in \insetof{\vcurr})$. We require the following additional structural invariant. 
\begin{enumerate}[resume*]
  \item \label{inv-sk5} For all nodes $n$, if there exists an $i$ such that $\markof{n, i}$ is set to $\false$, then $\markof{n, 0}$ is set to $\false$. 
\end{enumerate}
Invariant~\ref{inv-sk5} captures the fact that the base level gets marked at the end. This fact is useful for us because we can combine it with $\past (\markof{\vpred, i} = \false)$ to obtain $\past (\markof{\vpred, 0} = \false)$. Since \code{findNext} has found $\vpred$ to be marked at the base level, we can determine that it got marked at some time point between the point in time when $\past (\markof{\vpred, 0} = \false)$ held and the present. Let $t$ be the time point right before $\vpred$ got marked. The following invariant gives us useful information about which node $\nextof{\vpred, 0}$ can be pointing to at time $t$:
\begin{enumerate}[resume*]
  \item \label{inv-sk6}  For all nodes $n$ and level $i$, once $\markof{n, i}$ is set to $\true$, $\nextof{n, i}$ does not change.
\end{enumerate}
Invariant~\ref{inv-sk6} preceisely says that $\nextof{\vpred, 0} = \vcurr$ at time $t$. Thus, we have found a moment in time when $\markof{\vpred} = \false$, $\keyof{\vpred} < k$ and $\nextof{\vpred, 0} = \vcurr$. These three facts allow us to immediately conclude that $k \in \insetof{\vcurr}$ at time $t$. This finally completes the proof of \code{traverse}. 

\subsection{Lazy Traverse}
\label{sec-lazy-traverse}

\newcommand{\lazyRec}{\code{lazy\_rec}}
\newcommand{\lazyI}{\code{lazy\_i}}

The code for the lazy traversal is shown in \Cref{fig:lazy-traversal}. While the $\lazyRec$ implementation is almost identical to $\eagerRec$, the two differ in the per-level traversal in $\lazyI$ and $\eagerI$. The $\lazyI$ function here keeps track of three nodes while traversing: node $\vpred$ is the last unmarked node it witnessed, node $\vprednext$ is the node that $\vpred$ was pointing to when $\vpred$ was traversed, and $\vcurr$ is the node that is currently being traversed. The function $\lazyI$ begins by reading the next pointer and mark bit of $\vcurr$. If $\vcurr$ is found to be marked, then $\lazyI$ simply continues with the successor of $\vcurr$. If $\vcurr$ is unmarked, then the operation key $k$ is compared with $\keyof{\vcurr}$. In case $\keyof{\vcurr} < k$, then again $\lazyI$ continues with $\vcurr$ as the last seen unmarked node. Otherwise, $k \le \keyof{\vcurr}$. In this case, it attempts to unlink the marked segment between nodes $\vpred$ and $\vcurr$. The conditional on Line~\ref{lazyI-if} checks if there is a segment to remove. If that is indeed the case, then $\lazyI$ calls \code{changeNext} to unlink the segment between $\vpred$ and $\vcurr$. If \code{changeNext} succeeds, then there is a further check to make sure $\vcurr$ is still unmarked. If so, then $\lazyI$ returns with $\vpred$ and $\vcurr$. In all other cases where \code{changeNext} fails or $\vcurr$ is found to be marked, $\lazyI$ restarts the \code{traverse} function. 
 
\begin{figure}[h]
  \centering
  \begin{minipage}[t]{.60\textwidth}
\begin{lstlisting}[aboveskip=0pt,belowskip=0pt, frame=none]
let $\lazyI$ $i$ $k$ $\vpred$ $\vprednext$ $\vcurr$ =   
  match |<findNext>| $i$ $\vcurr$ with
  | $\vcurrnext$, $\true$ -> $\lazyI$ $i$ $k$ $\vpred$ $\vprednext$ $\vcurrnext$@\label{lazyI-rec1}@
  | $\vcurrnext$, $\false$ ->
    let $\vcurrk$ = |<getKey>| $\vcurr$ in
    if $kc$ < $k$ then
      $\lazyI$ $i$ $k$ $\vcurr$ $\vcurrnext$ $\vcurrnext$@\label{lazyI-rec2}@
    else
      let $\vres$ = ($kc$ = $k$ ? $\true$ : $\false$) in
      if $\vprednext$ = $\vcurr$ then@\label{lazyI-if}@
        ($\vpred$, $\vcurr$, $\vres$)@\label{lazyI-ret1}@
      else
        match |<changeNext>| $i$ $\vpred$ $\vprednext$ $\vcurr$ with
        | Success ->
          let _, $b$ = |<findNext>| $i$ $\vcurr$ in
          if $b$ then traverse $\vparr$ $\vsarr$ $k$
          else ($\vpred$, $\vcurr$, $\vres$)@\label{lazyI-ret2}@
        | Failure -> traverse $\vparr$ $\vsarr$ $k$
\end{lstlisting}
\end{minipage}%
\begin{minipage}[t]{.50\textwidth}
\begin{lstlisting}[aboveskip=0pt,belowskip=0pt, frame=none, firstnumber=last]
let $\lazyRec$ $i$ $\vparr$ $\vsarr$ $k$ =
  let $\vpred$ = $\vparr$[$i$+1] in
  let $\vcurr$, _ = |<findNext>| $i$ $\vpred$ in
  let $\vpredp$, $\vcurrp$, $\vres$ = $\lazyI$ $i$ $k$ $\vpred$ $\vcurr$ $\vcurr$ in
  $\vparr$[$i$] <- $\vpredp$;
  $\vsarr$[$i$] <- $\vcurrp$;
  if $i$ = 0 then
    ($\vpredp$, $\vcurrp$, $\vres$)
  else
    $\lazyRec$ ($i$-1) $\vparr$ $\vsarr$ $k$

let traverse $\vparr$ $\vsarr$ $k$ =
  $\lazyRec$ ($\vL$ - 2) $\vparr$ $\vsarr$ $k$
\end{lstlisting}
\end{minipage}
\caption{\label{fig:lazy-traversal} The lazy traversal for the skiplist template corresponding to the Harris List style traversal. }
\end{figure}

\subsection{Proof of Lazy Traversal}
\label{sec-lazy-proof}

The proof idea for the lazy traversal is similar to that of the eager traversal.  We provide a brief overview of the traversal invariant for $\lazyI$, and why it is sufficient to establish the required specification for \code{traverse}. The following discussion focuses on the $\lazyI$ executing on the base level, for sake of simplicity.  

\newcommand{\lazyinv}{\mathsf{LazyInv}}
\newcommand{\markedseg}{\mathsf{MarkedSeg}}
\newcommand{\vls}{\mathit{ls}}

The traversal invariant for $\lazyI$ is defined as below:   
\begin{align*}
  \lazyinv(k, \vpred, \vprednext, \vcurr) \defeq {} & \eagerinv(k, \vpred, \vprednext) * (\vprednext \neq \vcurr \impl \exists\; \vls.\; \markedseg(\vprednext, \vls, \vcurr))\\
  \markedseg(\vprednext, \vls, \vcurr) \defeq {} & (\vls = \texttt{[]} \impl \past (\nextof{\vprednext, i} = \vcurr))\\
  {} & * (\vls = \texttt{[\ensuremath{n_0},...,\ensuremath{n_l}]} \impl \past (\nextof{\vprednext, 0} = n_0) \\
  {} & \hspace{9.5em} * (\forall\; 0 \le j < l.\; \past \nextof{n_j, 0} = n_{j+1}))\\
  {} & \hspace{9.5em} * \past (\nextof{n_l, 0} = \vcurr)
\end{align*}
The invariant $\lazyinv$ relies on $\eagerinv$ and a new predicate $\markedseg$ that stores the segment of marked nodes. Additionally, the predicate establishes that the nodes $\vprednext$ and $\vcurr$ are the start and end points of the marked segment.

To see that $\lazyinv$ is indeed an invariant, consider Lines~\ref{lazyI-rec1} and \ref{lazyI-rec2} where $\lazyI$ is recursively called. At Line~\ref{lazyI-rec1}, the marked segment can be extended by including the marked node $\vcurr$, while $\vpred$ and $\vprednext$ remain unchanged. On the other hand, at Line~\ref{lazyI-rec2}, the second conjunct of $\lazyinv$ becomes trivially true, and $\
eagerinv(k, \vcurr, \vcurrnext)$ can be established in the same way as for $\eagerI$.

Let us now show why $\lazyinv$ is sufficient to prove the \code{traverse} specification. Consider the points where $\lazyI$ terminates (lines~\ref{lazyI-ret1} and \ref{lazyI-ret2}). When terminating at Line~\ref{lazyI-ret1}, we know that $\vprednext = \vcurr$. Hence, $\lazyinv$ provides the predicate $\eagerinv(k, \vpred, \vcurr)$. Using the same argument as $\eagerI$, we can establish the required postcondition. Now consider Line~\ref{lazyI-ret2}. In this case, the thread has successfully unlinked the marked segment, hence it can establish that $\markof{\vpred, 0} = \false$ and $\nextof{\vpred, 0} = \vcurr$ right after the call to \code{changeNext}. Note that the later check whether $\vcurr$ is unmarked must succeed in order for $\lazyI$ to terminate. Hence, at the point of the successful \code{changeNext}, $\vcurr$ must also be unmarked, giving us all the facts necessary to establish the required postcondition. This completes the proof of $\lazyI$.


\section{Soundness of Hindsight Specification}
\label{sec-hindsight-proof}

This section focuses on the soundness of the Hindsight specification, i.e., the proof that any structure that satisfies $(\ref{eq:hindsight-spec})$ must also satisfy $(\ref{eq:client-spec})$. We begin with some background on Iris that will be necessary for the proof, followed by additional details on the predicate $\historyinv$ and the helping invariant $\helpinginv$. Finally, we provide a proof outline relating $(\ref{eq:hindsight-spec})$ to the $(\ref{eq:client-spec})$.

\subsection{Iris Primer}
\label{sec-iris-primer}

Iris is a state-based concurrent separation logic. Assertions in the logic express ownership of resources. These resources include both physical resources that are manipulated by the program as well as ghost resources that capture auxiliary information for the purpose of the proof and that are manipulated using the proof rules of the logic. An example of an assertion describing a physical resource is the points-to predicate, $\ell \mapsto v$, which expresses ownership of heap location $\ell$ and constrains its stored value to be $v$. Two assertions $P$ and $Q$ (e.g. describing disjoint heap regions) can be composed using a\emph{separating conjunction}, $P * Q$. The logic supports standard local reasoning principles such as the \emph{frame rule}: if the Hoare triple $\hoareTriple{P}{e}{Q}$ is valid, then so is $\hoareTriple{P * R}{e}{Q * R}$ for all \emph{frames} $R$.

To describe ghost resources, Iris provides assertions of the form $\ghostState{}{\melt}$ which state ownership of a chunk $\melt$ of the ghost resource stored at ghost location $\gamma$. The chunk $\melt$ must be drawn from a \emph{camera} that is specified as part of the proof. Cameras generalize the partial commutative monoids that are commonly used to define the semantics of separating conjunction. The precise definition of these structures is immaterial for this paper. What is important is that a camera is a set $S$ equipped with an operation $\melt \cdot \meltB$ that is compatible with separating conjunction in the following way: $\ghostState{}{\melt} * \ghostState{}{\meltB} \provesIff \ghostState{}{\melt \raOp \meltB}$.
\newcommand{\maxnat}{\mathsf{Max}}
A simple example is the $\mathsf{MaxNat}$ camera defined by $S = \setcomp{\maxnat(n)}{n \in \Nat}$ and $\maxnat(n)\cdot\maxnat(m)=\maxnat(\max(n,m))$. 

A proof can introduce ghost resources as needed by allocating a fresh ghost location $\gamma$. A proof can also update chunks of ghost resources at existing ghost locations using \emph{view shifts}, $\ghostState{}{\melt} \vsR \ghostState{}{\meltB}$. A view shift is only allowed if replacing $\melt$ by $\meltB$ is \emph{frame-preserving}. Intuitively, this means that for any $\meltC$ such that the full ghost resource stored at $\gamma$ may be $\melt \cdot \meltC$ before the update, $\meltB$ must also compose with $\meltC$. For example, in the $\mathsf{MaxNat}$ camera, all updates are frame-preserving because composition (maximum) is totally defined. However, in general, this is not the case.


\newcommand{\agree}{\mathsf{agree}}
\newcommand{\Ag}{\mathsf{Ag}}
\newcommand{\Map}{\mathsf{Map}}

To reduce the burden on the proof author, Iris provides a library of predefined parameterized cameras. One such camera that we will use below is the \emph{agreement camera} $\Ag(X)$ which is defined for any set $X$. The elements take the form $\agree(x)$ for $x \in X$ and composition is only defined in the case of identity: $\agree(x)\cdot\agree(x)=\agree(x)$. Another predefined camera we will use is that of finite partial maps from some set $X$ to some camera $R$, $\Map(X,R)$, with composition defined by lifting the composition on $R$ pointwise to maps.

To enable reasoning about shared resources, Iris provides the notion of an \emph{invariant}. Any Iris assertion $P$ can be turned into an invariant, denoted $\sinv{P}$. Once an invariant has been created, it becomes a duplicable resource, which means that it can be commonly owned by all threads. An individual thread may access (and manipulate) the resources described by the underlying assertion $P$ when performing a computation step, provided $P$ is reestablished after each step.

A reoccurring pattern in Iris proofs is to have an invariant that holds authoritative ownership of the current state of a resource, and each thread may additionally own partial knowledge about this resource that the thread gathered during its execution.
To encode this pattern, Iris provides the \emph{authoritative camera} $\mathsf{Auth(R)}$ defined over any camera $R$.
The authoritative camera has elements of the form $\authAuth\; a$ to denote authoritative ownership of $a \in R$ and $\authFrag\; a$ to denote \emph{fragmental} ownership. The composition $\authAuth\;a \raOp \authFrag\;b$ expresses the fact that $b$ is a fragment of $a$, i.e., $\exists\; c, a = b \raOp c$. For example, the camera $\mathsf{Auth(MaxNat)}$ is very useful to establish lower bounds of a monotonically non-decreasing quantity. The definitions of $\mathsf{Auth(MaxNat)}$ yields the valid entailment: $\ghostState{}{\authAuth\; \maxnat(n)} * \ghostState{}{\authFrag\; \maxnat(n')} \vdash n' \leq n$. The permitted frame-preserving updates are shown below:
\begin{mathpar}
  \inferH{auth-max-upd}{n \leq n'}{\ghostState{}{\authAuth \maxnat(n)} \vsR \ghostState{}{\authAuth \maxnat(n')}}

  \inferH{auth-max-snap}{}
  {\ghostState{}{\authAuth \maxnat(n)} \vsR \ghostState{}{\authAuth \maxnat(n)} * \ghostState{}{\authFrag \maxnat(n)}}
%
\end{mathpar}
The rule \refRule{auth-max-upd} requires that the update only increases the authoritative value to guarantee that the new authoritative value $n'$ still composes with all fragmental values.

\subsection{Tracking the Computation History}
\label{sec-hindsight-history}

Hindsight arguments involve reasoning about past program states. Our encoding therefore tracks information about past states using \emph{computation histories}. 
We define computation histories as finite partial maps from \emph{timestamps}, $\Nat$, to \emph{snapshots}, $\Snap$. A snapshot describes an abstract view of a program state at a particular time. It is a parameter of our encoding. For instance, a snapshot may capture the physical memory representation of the data structure under proof, while abstracting from the remainder of the program state.
Another parameter of our encoding is a function $\cssabs{\cdot}$ that computes the abstract state of the data structure from a given snapshot. 
 

The predicate $\historyinv(H, T)$ consists of two ghost resources: one resource to store the authoritative computation history $H$, and one resource to store the last timestamp $T$ used in $H$. Whenever the program changes the program state in a way that is observable at the level of snapshots, the invariant forces $H$ to be updated with a new snapshot at time $T+1$, and $T$ to be increased by $1$. To track $T$, we use the camera $\mathsf{Auth(MaxNat)}$, which implicitly encodes the fact that timestamps are non-decreasing, effectively serializing the snapshot changes in the history.

We want threads to be able to own fragmental knowledge of the authoritative history $H$. These history fragments are then used to define past predicates that express the existence of a past state with a certain snapshot. To this end, we use the camera $\Auth(\Map(\Nat,\Ag(\Snap)))$ for the resource that tracks the history. The use of the agreement camera ensures that the history can only be updated by appending to it (i.e., all snapshots of past states are preserved).

The predicate $\historyinv(H, T)$ is then defined as
\begin{align*}
    \historyinv(H, T) \defeq {} & \ghostState{t}{\authAuth\; \maxnat(T)} \,*\, \ghostState{m}{\authAuth\; \agree(H)} \\ 
    & {} \,*\, \dom(H) = \{0 \dots T\} \,*\,  \forall t < T.\, H(t) \neq H(t+1)\enspace.
\end{align*}
Here, we abuse notation and write $\agree(H)$ for the function that takes the history $H$ to an element of $\Auth(\Map(\Nat,\Ag(\Snap)))$ in the expected way. The second last conjunct ensures that the authoritative history contains no gaps. The purpose of the last conjunct is to ensure that a thread appends to the history whenever any change to the program state is made that affects the snapshot (see discussion in \cref{sec-hindsight-invariant}). 

\subsection{Augmenting \code{op} with Prophecies}
\label{sec-op-prophecies}

In light of the discussion in \cref{sec-hindsight-spec}, we must augment $\cssOp$ with auxiliary ghost code that creates and resolves the relevant prophecies. We do this by defining the wrapper function $\csslop$ given in \cref{fig-op'-algorithm}.
Let us first briefly discuss how prophecy variables can be used in Iris. The right side of \cref{fig-op'-algorithm} shows the specifications of the two functions related to manipulating prophecies.
The function $\newproph$ returns a fresh prophecy $p$ that predicts a sequence of values $pvs$, captured by the resource $\proph(p, \vpvs)$.
This resource can be owned and shared among threads via a shared invariant. The resource is also non-duplicable.

The values contained in the sequence $\vpvs$ are determined by how $p$ is resolved using the $\code{Resolve}$ command. The rule~\ruleref{prophecy-resolution} ties the observed value ($v$) to the next prophesied value (the head of $\vpvs$). It also updates $\proph(p, \vpvs)$ to the tail of $\vpvs$ for the remaining prophesied values that are yet to be observed. In our context, we want the prophecy to predict whether the decisive operations (i.e. \code{markNode} and \code{changeNext}) will succeed as well as their return value. Hence, we augment \code{delete} and \code{insert} by wrapping their decisive operations inside a \code{Resolve} command. The return value is captured by the resolution on Line~\ref{line-op'-resolve}. For our purposes, we assume that $\vpvs$ is a sequence of \code{Success} or \code{Failure} values (i.e., the result of each attempted call to the decisive operation) followed by a special value $\prophend{\vres}$ indicating that no further future calls are expected, and that $\vres$ will be the final return value\footnote{We have to also consider the case where $\vpvs$ is not of this form, but that is a  trivial case to handle and thus we do not consider it in this discussion.}. Given the prophesied sequence of values $\vpvs$, a thread will exhibit an unmodifying linearization point iff $\vpvs$ does not contain any \code{Success} values.

Let us now turn to $\csslop$, which starts by generating a unique thread identifier using $\newthreadid$ on Line~\ref{line-op'-tid}. We also implement $\newthreadid$ using the $\newproph$ construct. There are two benefits to this: (i) Iris makes sure that prophecy identifiers are unique out of the box, and (ii) now we can use Iris's erasure theorem for prophecies to argue that the augmented code $\csslop$ has the same behavior as $\cssOp$. After, $\csslop$ creates a prophecy $p$ as described above, it invokes the underlying operation $\cssOp$, and finally, terminates after resolving $p$ to the result of $\cssOp$.

\begin{figure}[t]
  \centering
\begin{minipage}[t]{.48\textwidth}
\begin{lstlisting}[aboveskip=0pt,belowskip=0pt, frame=none]
let $\csslop$ $r$ $k$ =
  let $\tid$ = $\newthreadid$ in @\label{line-op'-tid}@
  let $p$ = $\newproph$ in @\label{line-op'-proph}@
  let $v$ = $\cssOp$ $p$ $r$ $k$ in @\label{line-op'-op}@
  $\resolve{p}{\prophend{v}}$; $v$ @\label{line-op'-resolve}@
\end{lstlisting}
\end{minipage}\hfill
\begin{minipage}[t]{.48\textwidth}
  \strut\vspace*{-\baselineskip}\newline
\begin{mathpar}
  \inferHtop{prophecy-creation}
  {}
  {\hoareTriple{\mathsf{True}}{\newproph}{p.\; \exists\; \vpvs.\; \proph(p, \vpvs)}}
  \\
  \inferHtop{prophecy-resolution}
  {}
  {\annot{\proph(p, \vpvs)} \; \resolve{p}{v} \; \qquad\qquad \\ \annot{\exists\; \vpvss.\; \proph(p, \vpvss) \;*\; \vpvs = v :: \vpvss}}
\end{mathpar}
\end{minipage}
\caption{Wrapper augmenting $\cssOp$ with prophecy-related ghost code, whose specification is given on the right.
\label{fig-op'-algorithm}}
\end{figure}
We can now finally present the client-level specification with $\csslop$:
$$ \minvr \magicwand \atomicTriple{C.\; \cssstate(r, C)}{\csslop\;p\;r\;k}{\vres \; C'. \; \cssstate(r, C') \,*\, \dSpec{op}(k, C, C', res) }.$$ 
The next section provides details on our choice of $\mainInv(r)$ and the proof of $\csslop$. 

\subsection{Proof of $\csslop$}
\label{sec-op-proof}

\begin{figure}[t]

  \begin{align*}
    \minvr \defeq {} & \exists\, H\, T\, C.\; \csslstate(r, C) * \cssabs{H(T)} = C\\
    {} & * \historyinv(H, T)  \,*\, \helpinginv(H, T) \,*\, \tplinv(H, T) \\
    \helpinginv(H, T) \defeq {} & \exists\, R.\; \ghostState{r}{\authAuth\; R} \\
    {} & * \Sep_{\tid \in R}        
    \; \exists\, \cssOp\, k\, v_p\, t_0\, \Phi\, \token.\; \reg(\tid, \cssOp, k, v_p, t_0, \Phi, \token) \\
    \reg(\tid, \cssOp, k, v_p, t_0, \Phi, \token) \defeq {} &
    \thread(\tid, t_0) * \sinv{\helpingstate(\cssOp, k, v_p, t_0, \Phi, \token)}
    \\[.5em]
    \helpingstate(\cssOp, k, v_p, t_0, \Phi, \token) \defeq {}
    & \linpending(\cssOp, k, v_p, t_0, \Phi) \lor \lindone(\cssOp, k, v_p, t_0, \Phi, \token)
    \\[.5em]
    \linpending(\cssOp, k, v_p, t_0, \Phi) \defeq {}
    & 
    {\atomicUpdate}_{\cssOp}(\Phi) * \neg \pastlin(\cssOp, k, v_p, t_0)
    \\[.5em]
    \lindone(\cssOp, k, v_p, t_0, \Phi, \token) \defeq {}
    &  (\Phi(v_p)\; \lor\; \token) * \pastlin(\cssOp, k, v_p, t_0)
  \end{align*}
  \caption{Definition of the invariant encoding the helping protocol.}
  \label{fig-helping-prot-inv'}
\end{figure}

\refFig{fig-helping-prot-inv'} shows a simplified definition of the invariant that encodes the helping protocol\footnote{For presentation purposes, the proof outline presented here abstracts from some technical details of the actual proof in Iris.}.
The definitions are implicitly parameterized by a proposition $\tplinv(r,H,T)$, which abstracts from the resources needed for proving that a specific structure satisfies the hindsight specification.

The helping protocol predicate $\helpinginv(H, T)$ contains a \emph{registry} $\ghostState{r}{\authAuth\; R}$ of thread IDs with unmodifying linearization points that require helping from other concurrent threads.
For each thread ID $\tid$ in the registry, the shared state contains $\thread(\tid,\_)$ along with the state of $\tid$, which is either $\linpending$ or $\lindone$. $\linpending$ captures an uncommitted atomic triple, and $\lindone$ describes the operation after it has been committed. Note that we use the notation ${\atomicUpdate}_{\cssOp}(\Phi)$ to refer to the atomic update token of $\cssOp$ and write just ${\atomicUpdate}(\phi)$ when $\cssOp$ is clear.

\begin{figure}[]
  \centering
  \begin{lstlisting}[aboveskip=1pt,belowskip=0pt, frame=none]
$\minvr * \annotAtom{C.\; \cssstate(r,C)}$
let $\csslop$ $r$ $k$ =
  $\annot{\atomicUpdate(\Phi)}$ @\label{line-op-proof-logatom-intro}@
  let $\tid$ = $\newthreadid$
  $\annot{\atomicUpdate(\Phi) * \proph(\tid, \_)}$
  let $p$ = $\newproph$ in
  $\annot{\atomicUpdate(\Phi) * \proph(\tid, \_) * \proph(p, \vpvs) * \minvr}$ @\label{line-op'-proof-case-split}@ 
  (* Open invariant *)
  $\annot{\atomicUpdate(\Phi) * \thread(\tid, T_0) * \proph(p, \vpvs) * \csslstate(r, C_0) * \helpinginv(H_0, T_0) * \dots}$ 
  (* Case analysis on $\procProph(\vpvs)$ : only showing $\neg \procProph(\vpvs)$ *)
  $\annot{\atomicUpdate(\Phi) * \thread(\tid, T_0) * \proph(p, \vpvs) * \csslstate(r, C_0) * \ghostState{r}{\authAuth\; R} * \tid \notin R * \dots}$ 
  (* Let $v_p$ such that $\vpvs$ contains $\prophend{v_p}$ *)
  (* Case analysis on $\dSpec{op}(k, \abs{H_0(T_0)}, \abs{H_0(T_0)}, v_p)$ : only showing $\neg \dSpec{op}(\abs{H_0(T_0)}, \abs{H_0(T_0)}, v_p)$ *)
  $\annot{\thread(\tid, T_0) * \proph(p, \vpvs) * \csslstate(r, C_0) * \ghostState{r}{\authAuth\; R} * \tid \notin R}$ @\label{line-op'-proof-helping-case}@
    $\annot{\dots * \atomicUpdate(\Phi) * \thread(\tid, T_0) * \neg \pastlin(\cssOp, k, v_p, T_0)}$
    $\annot{\dots * \atomicUpdate(\Phi) * \thread(\tid, T_0) * \linpending(\cssOp, k, v_p, T_0, \Phi) * \token}$
    $\annot{\dots * \helpingstate(\cssOp, k, v_p, T_0, \Phi, \token) * \token}$
  (* Ghost update: $\ghostState{r}{\authAuth\; R} \vsR \ghostState{r}{\authAuth\; R \cup \set{\tid}}$ *) @\label{line-op'-proof-helping-register}@
  $\annot{\thread(\tid, T_0) * \proph(p, \vpvs) * \ghostState{r}{\authFrag\; \{tid\}} * \token * \csslstate(r, C_0) * \ghostState{r}{\authAuth\; R \cup \{tid\}} * \dots }$
  $\annot{\thread(\tid, T_0) * \proph(p, \vpvs) * \ghostState{r}{\authFrag\; \{tid\}} * \token * \csslstate(r, C_0) * \helpinginv(H_0, T_0) * \dots }$
  (* Close invariant *)
  $\annot{\thread(\tid, T_0) * \proph(p, \vpvs) * \ghostState{r}{\authFrag\; \{tid\}} * \token }$
  let $v$ = $\cssOp$ $p$ $r$ $k$ in @\label{line-op'-proof-call-op}@
  $\annot{\thread(\tid, T_0) * \ghostState{r}{\authFrag\; \{tid\}} * \token * \proph(p, \vpvss) * (\vpvs = \vprf \;\code{++}\; \vpvss) \\ * (\procProph(\vpvs) \magicwand \Phi(v)) * (\neg \procProph(\vpvs) \magicwand \pastlin(\cssOp, k, v_p, T_0))}$ @\label{line-op'-proof-op-post}@
  $\annot{\thread(\tid, T_0) * \ghostState{r}{\authFrag\; \{tid\}} * \token * \proph(p, \vpvss) * \pastlin(\cssOp, k, v_p, T_0)}$
  $\resolve{p}{\prophend{v}}$; @\label{line-op'-proof-resolve-p}@
  $\annot{\thread(\tid, T_0) * \ghostState{r}{\authFrag\; \{tid\}} * \token * \proph(p, \code{[]}) * (v_p = v) * \pastlin(\cssOp, k, v_p, T_0)}$
  (* Open invariant *)
  $\annot{\thread(\tid, T_0) * \ghostState{r}{\authFrag\; \{tid\}} * \token * \pastlin(\cssOp, k, v, T_0) * \cssstate(r, C_1) * \helpinginv(H_1, T_1) * \dots}$
  $\annot{\dots * \token * \ghostState{r}{\authFrag\; \{\tid\}} * \pastlin(\cssOp, k, v, T_0) * \helpingstate(\cssOp, k, v, T_0, \Phi, \token)}$
  $\annot{\dots * \token * \pastlin(\cssOp, k, v, T_0) * \lindone(\cssOp, k, v, T_0, \Phi, \token)}$
  $\annot{\dots * \token * \pastlin(\cssOp, k, v, T_0) * (\Phi(v) \lor \token)}$
  $\annot{\dots * \Phi(v) * \pastlin(\cssOp, k, v, T_0) * (\Phi(v) \lor \token) }$
  $\annot{\dots * \Phi(v) * \pastlin(\cssOp, k, v, T_0) * \lindone(\cssOp, k, v, T_0, \Phi, \token)}$
  $\annot{\Phi(v) * \cssstate(r,C_1) * \helpinginv(H_1, T_1) * \dots}$
  (* Close invariant *)
  $\annot{\Phi(v)}$
  $v$
  $\annotAtom{\vres \; C'. \; \cssstate(r, C') \,*\, \dSpec{op}(C, C', res)}$
\end{lstlisting}

\caption{Outline for the proof of the client-level specification for \code{op}.}
\label{fig-op'-proof}
\end{figure}

The proof outline for $\csslop$ is shown in \refFig{fig-op'-proof}. The proof begins by creating a thread identifier $\vtid$ and prophecy $p$. As alluded to earlier, $\newthreadid$ is also just $\newproph$ in disguise, hence we obtain two prophecy resources, one each for $\vtid$ and $p$. Before we can invoke $\cssOp$, we must record the start time of this thread and perform a case split on $\procProph(\vpvs)$ to determine if this thread requires helping. For this, we open the invariant at Line~\ref{line-op'-proof-case-split}. To record the start time, we give up the resource $\proph(\vtid, \_)$ to gain a duplicable resource $\thread(\vtid, T_0)$ in exchange ( $T_0$ is the current timestamp). The exact details of this exchange are not relevant, so we continue with the case split on $\procProph(\vpvs)$. First consider the case that $\procProph(\vpvs)$ holds. Then, thread $\vtid$ does not require helping and the hindsight specification can ensures that $\cssOp$ linearizes correctly. This is quite straightforward, so we focus on the other case, i.e. $\procProph(\vpvs)$ does not hold.

In order to register the thread for helping, we must establish $\pastlin(\cssOp, k, v_p, T_0)$ or its negation. Here, $v_p$ is the return value obtained by scanning for the end marker in $\vpvs$. We proceed by first checking if the thread can be linearized at the current timestamp $T_0$. If so, then we can safely linearize the thread right away. So let us focus on the more interesting case where we cannot linearize right away (Line~\ref{line-op'-proof-helping-case}). Here, the thread registers itself with the helping protocol by updating $R$ to $R \cup \set{\vtid}$ using an update rule for the authoritative set camera (in Line~\ref{line-op'-proof-helping-register}). As part of registering for helping, it first establishes $\linpending(\cssOp, k, v_p, T_0, \Phi)$ by transferring its obligation to linearize to the shared state, captured by the update token $\atomicUpdate(\Phi)$. The condition $\pastlin(\cssOp, k, v_p, T_0)$ follows from the fact that $\dSpec{op}(k, \abs{H_0(T_0)}, \abs{H_0(T_0)}, v_p)$ does not hold. The thread also creates a fresh non-duplicable token $\token$ that it will later trade in for the receipt $\Phi(v_p)$.

We are now at the point of invocation for $\cssOp$. Before continuing further, let us briefly switch to the role played by the concurrent thread that must linearize thread $\tid$. The helping thread must update the structure, and as per the invariant, update the history $M$ as well as the helping protocol $\helpinginv(H,T)$. In particular, while updating the helping protocol, it scans over all threads registered for helping so far, moving them from state $\linpending$ to state $\lindone$ as per the prophesied return value $v_p$.

Let us now return to the proof thread $\vtid$ at Line~\ref{line-op'-proof-call-op}. As precondition of $(\ref{eq:hindsight-spec})$, we give away $\proph(p, \vpvs)$ and obtain the postcondition (Line~\ref{line-op'-proof-op-post}). Simplifying the postcondition of $(\ref{eq:hindsight-spec})$ for our case $\neg \procProph(\vpvs)$, we obtain the predicate $\pastlin(\cssOp, k, v_p, T_0)$.
Next, we make a final resolution of the prophecy on Line~\ref{line-op'-proof-resolve-p}. Since, $\vpvss$ is the suffix of $\vpvs$, $\prophend{v_p}$ must also be the final item in $\vpvss$. After resolution, we obtain $\prophend{v_p} = \prophend{v}$ and by injectivity of the end marker, $v_p = v$. 

We now have the predicate $\pastlin(\cssOp, k, v, T_0)$ which we use to obtain the receipt of linearization from the shared invariant. Because $\pastlin(\cssOp, k, v, T_0)$ holds now, we know thread $\vtid$ cannot be in the $\linpending$ state. Hence, we know that thread $\vtid$ must be in the $\lindone$ state. Since the thread owns the unique token $\token$, it trades it in to obtain $\Phi(v)$, which lets it complete the proof of its client-level specification.


\section{Additional detail on verification of the Skiplist Template}
\label{sec-template-proof-extra}

\subsection{Keysets Reasoning}

Recall that keysets exhibit two properties that are crucial for localizing the functional correctness specification of search structures from the global data structure graph to individual nodes in the graph: (1) keysets partition the set of all keys, and (2) a node's contents is always a subset of its keyset. To capture these properties naturally in Iris, \cite{DBLP:conf/pldi/KrishnaPSW20} introduces a \emph{keyset camera}. An element of the keyset camera is a pair of sets of keys $(K, C)$ with $C \subseteq K$. We associate such pairs to regions in the data structure graph such that $C$ is the union of the node contents in the region and $K$ is the union of the nodes keysets. The composition of such pairs is defined as (pair-wise) disjoint union. The validity and the composition operator of the keyset camera then implicitly capture the desired properties (1) and (2) above.

Using this camera we tie the keyset reasoning to the invariant for the skiplist template by instantiating the predicate $\resourcesks$ (from \cref{sec-template-proof}) as follows:
\[\resourcesks(s) \defeq  \ghostState{}{\authAuth (\KS, C(s))} * \Sep_{n \in \fpof{s}} \ghostState{}{\authFrag (\keysetof{s, n}, \contentsof{s, n})}\enspace.\]
The assertion $\ghostState{}{\authAuth (\KS, C(s))}$ signifies ownership of the logical contents $C(s)$ of the structure in the state captured by snapshot $s$. Recall that $\KS$ is the set of all keys. The set $\fpof{s}$ tracks the \emph{footprint} of $s$, i.e., the set of all nodes in the data structure graph captured by $s$. Each individual node $n \in \fpof{s}$ owns its keyset $\keysetof{s,n}$ and node-local logical contents $\contentsof{s,n}$. The keyset camera enforces that the $\keysetof{s,n}$ are pairwise disjoint, $\contentsof{s,n} \subseteq \keysetof{s,n}$ for each $n$, and that $C(s) = \bigcup_{n \in \fpof{s}} \contentsof{s,n}$, respectively,
$\keyspace = \bigcup_{n \in \fpof{s}} \keysetof{s,n}$.

A node's keyset and contents are then tied to the actual physical representation of the data structure via an abstract predicate that is instantiated for each node implementation. Recall that a node's keyset is defined in terms of its inset, an inductive quantity defined over the edgeset graph of the structure. As in~\cite{DBLP:conf/pldi/KrishnaPSW20}, we use the Flow Framework~\cite{DBLP:journals/pacmpl/KrishnaSW18,DBLP:conf/esop/KrishnaSW20} to obtain a node-local characterization of the inset quantity and to reason about insets in Iris. We omit these details here as they are not relevant for understanding the high-level proof idea.

\subsection{Helper Function Specifications}

\newcommand{\prophcolor}[1]{\textcolor{brown}{#1}}

Our encoding of the hindsight specification $(\ref{eq:hindsight-spec})$ requires us to resolve a prophecy around each program step that may modify the structure. In the context of the skiplist template, the structure is modified using calls to the helper functions \code{markNode} and \code{changeNext} on the base-level list. 
Unfortunately, prophecies in Iris can only be resolved around physical computation steps in the Iris programming language HeapLang and not at a more abstract level around logical atomic triples. Hence, the atomic triples for \code{markNode} and \code{changeNext} themselves must capture the relevant aspects of the reasoning related to prophecy resolution. 

\Cref{fig-template-helper-specs-extra} shows these atomic triple specifications.
We focus on the \code{markNode} specification in particular. The assertions in \textcolor{blue}{blue} are those already described in \cref{fig-template-helper-specs}. The additional assertions in \textcolor{brown}{brown} are concerned with the prophecy reasoning. The precondition $\prophcolor{\proph(p, \vpvs)}$ gives the right to resolve the prophecy $p$. The additional postcondition says that \code{marknode} resolves the prophecy an arbitrary number of times, but if it succeeds, then a \code{Success} value was seen in the list of prophesied values $\vpvs$, yielding $\procProph(\vpvs)$. Otherwise, it says that no \code{Success} value was seen so far, yielding $\procProph(\vprf)$. The template proof for \code{delete} then uses this information to infer whether $\procProph(\vpvs)$ holds and prove the appropriate postcondition from the hindsight specification. The specification of \code{changeNext} is similar.

\begin{figure}
  \centering
  \begin{lstlisting}
    $\annotAtom{k\,h\,\vmark\,\vnext.\; \nodeInv(n, k, h, \vmark, \vnext) * (0 < h) \prophcolor{* \proph(p, \vpvs)}}$ 
                        |<markNode>| $0$ $n$ $p$
$\annotAtom{\Ret x.\; \nodeInv(n, k, h, \vmarkp, \vnext) \prophcolor{* \proph(p, \vpvss) * (\vpvs = \vprf \;\code{++}\; \vpvss)}\\ * (\vmark(0) = \false \impl x = \mathsf{Success} * \vmarkp = \vmark[0 \rightarrowtail \true] \prophcolor{* \procProph(\vpvs)}) \\ * (\vmark(0) = \true \impl x = \mathsf{Failure} * \vmarkp = \vmark \prophcolor{* \neg \procProph(\vprf)})}$

    $\annotAtom{k\,h\,\vmark\,\vnext.\; \nodeInv(n, k, h, \vmark, \vnext) * (0 < h) \prophcolor{* \proph(p, \vpvs)}}$ 
                        |<changeNext>| $0$ $n$ $n'$ $e$ $p$
$\annotAtom{\Ret x.\; \nodeInv(n, k, h, \vmark, \vnextp) \prophcolor{* \proph(p, \vpvss) * (\vpvs = \vprf \;\code{++}\; \vpvss)} \\ * ((\vmark(0) = \false \land \vnext(0) = n') \impl x = \mathsf{Success} * \vmarkp = \vmark[0 \rightarrowtail \true] \prophcolor{* \procProph(\vpvs)}) \\ * ((\vmark(0) = \true \lor \vnext(0) \neq n') \impl x = \mathsf{Failure} * \vmarkp = \vmark \prophcolor{* \neg \procProph(\vprf)})}$
\end{lstlisting}
\caption{Specifications of the helper functions with the prophecies.}
\label{fig-template-helper-specs-extra}
\end{figure}

\subsection{Proof outline of \code{delete}}
\label{sec-template-proof-extra-delete}

In this section, we discuss the proof of \code{code} in more detail.
The proof outline is shown in \cref{fig-delete-proof}. We want to show that \code{delete} satisfies $(\ref{eq:hindsight-spec})$. The precondition provides access to the invariant $\mainInv(r)$ and knowledge that the thread ID is $\vtid$ with start time $t_0$. Additionally, the precondition provides the conjunct $(\procProph(\vpvs) \magicwand \atomicUpdate(\Phi))$, signifying that the thread can perform its linearization in case of a successful decisive operation. The precondition also provides the right to resolve prophecies. It is easy to see how prophecies are mainpulated by the helper functions from the specifications in \cref{fig-template-helper-specs-extra}, hence we ignore them in the proof outline. 

The algorithm of \code{delete} begins with a call to \code{traverse}. The specification for \code{traverse} provides the postcondition
\[ \pastt{s, t_0}(k \in \keysetof{s, \vcurr} \land (\vres \iff k \in \contentsof{s, \vcurr})).\]
Note that we use the past operator $\pastt{s, t_0}$ from \cref{sec-hindsight-invariant} in the postcondition of \code{traverse}. Intuitively, this assertion captures that there is a past state $s$ in the history (after time point $t_0$) in which $k$ is in the keyset of $\vcurr$ and $\vres$ is true iff $k$ is in the logical contents of $\vcurr$. Note that this implies $\markof{s, \vcurr, 0} = \false$ due to Invariant~\ref{inv-sk1}.

The \code{delete} algorithm proceeds by case analysis on the result of \code{traverse}. Let us first consider the case that $\vres$ is $\false$ (Line~\ref{delete-proof-res-false}). The \code{delete} operation terminates with result $\false$. The specification (\ref{eq:hindsight-spec}) requires establishing the predicate $\pastlin(\code{del}, k, \false, t_0)$ in this case. To this end, we must provide a witness past state in which $k$ was not part of the abstract state. We obtain this witness from the postcondition of \code{traverse} which provides a past state $s$ such that $k \in \keysetof{s, c}$ and $k \notin \contentsof{s ,c}$. Applying (\ref{eqn-keyset}) in state $s$, we can establish the predicate $\pastlin(\code{del}, k, \false, t_0)$ (Line~\ref{delete-proof-post1}).

Let us now consider the case when $\vres$ is $\true$. The \code{delete} algorithm here calls the helper function \code{maintainanceOp\_del}, which marks node $\vcurr$ at the higher levels. Thus, after \code{maintainanceOp\_del} terminates, we obtain the information that all higher levels of $\vcurr$ are marked (Line~\ref{delete-proof-maintenance-post}). Next, the helper function \code{markNode} is called at the base level to logically delete $\vcurr$ from the structure. In order to apply the specification of \code{markNode}, we open the invariant $\mainInv(r)$ to access the $\nodeInv$ predicate for $\vcurr$ (Line~\ref{delete-proof-marknode-pre}). Let the current snapshot at this point (obtained by opening the invariant) be $s_0$. The call to \code{markNode} call either succeeds or fails depending on whether $\markof{s_0, \vcurr, 0} = \false$ holds. We consider both of the cases in turn. 

First, suppose that \code{markNode} succeeds (Line~\ref{delete-proof-marknode-success}). This implies that $\markof{s_0, \vcurr, 0} = \false$ and a new snapshot $s_1$ must be constructed to record the marking of $\vcurr$. The proof then appends $s_1$ to the history stored in $\mainInv(r)$ in order to reestablish $\mainInv(r)$. Additionally, success of \code{markNode} implies that $\procProph(\vpvs)$ holds. Hence, the thread must use the resource $\atomicUpdate(\Phi)$ to linearize itself. The receipt of linearization $\Phi$ is exactly what is required in the postcondition of (\ref{eq:hindsight-spec}) in this case (Line~\ref{delete-proof-post2}).

Finally, consider the case that \code{markNode} fails (Line~\ref{delete-proof-marknode-failure}). The \code{delete} algorithm terminates with result $\false$ in this case. Hence, we must again establish $\pastlin(\code{del}, k, \false, t_0)$ to complete the proof. That is, we must identify a witness past state such that $k$ was not part of the abstract state of the structure. To compute this witness, note that $\markof{s, \vcurr, 0} = false$ for state $s$ observed by \code{traverse}, while $\markof{s_0, \vcurr, 0} = \true$ for the current state $s_0$. By $\transinv$ in \cref{fig-template-tplinv} and temporal interpolation, there must exist consecutive states $s_1$ and $s_2$ when $\vcurr$ was marked at the base level (Line~\ref{delete-proof-temporal-interpolation}). Here, the predicate $\paststate(s, t)$ represents the knowledge that the history recorded snapshot $s$ at time $t$. The snapshot $s_2$ is the required witness to establish the postcondition of $(\ref{eq:hindsight-spec})$ (Line~\ref{delete-proof-post3}). This completes the proof of \code{delete}.

\begin{figure}[]
  \centering
  \begin{lstlisting}[aboveskip=1pt,belowskip=0pt, frame=none]
$\annot{\minvr * \thread(\vtid, t_0) * (\procProph(\vpvs) \magicwand \atomicUpdate(\Phi))}$
let delete $k$ =
  let $\vparr$ = allocArr $\vL$ $\vhead$ in
  let $\vsarr$ = allocArr $\vL$ $\vtail$ in
  let $\vpred$, $\vcurr$, $\vres$ = traverse $\vparr$ $\vsarr$ $k$ in@\label{line-delete-traverse}@
  $\annot{\pastt{s, t_0}(k \in \keysetof{s, \vcurr} \land (\vres \iff k \in \contentsof{s, \vcurr}))}$
  if not $\vres$ then
    $\annot{\pastt{s, t_0}(k \in \keysetof{s, \vcurr} \land (k \notin \contentsof{s, \vcurr}))}$@\label{delete-proof-res-false}@
    $\annot{\pastt{s, t_0}(k \notin \abs{s})}$
    $\annot{\pastt{s, t_0}(\dSpec{del}(k, \abs{s}, \abs{s}, \false))}$
    $\false$
    $\annot{\neg \procProph(pvs) * \pastlin(\code{del}, k, \false, t_0)}$ @\label{delete-proof-post1}@
  else
    $\annot{\pastt{s, t_0}(k \in \keysetof{s, \vcurr} \land (k \in \contentsof{s, \vcurr}))}$
    maintainanceOp_del $\vcurr$; 
    $\annot{\pastt{s, t_0}(\markof{s, \vcurr, 0} = \false) * \pastt{s', t_0}(\forall i, 0 < i < \htof{s', \vcurr} \impl \markof{s', \vcurr, i} = \true)}$@\label{delete-proof-maintenance-post}@
    $\annot{\pastt{s, t_0}(\markof{s, \vcurr, 0} = \false) * \historyinv(H_0, T_0) * \resources(s_0) * \cdots}$
    $\annot{\pastt{s, t_0}(\markof{s, \vcurr, 0} = \false) * \historyinv(H_0, T_0) * \nodeInv(\vcurr, s_0) * \cdots}$@\label{delete-proof-marknode-pre}@
    match |<markNode>| 0 $\vcurr$ with
    | Success -> 
      $\annot{(\markof{s_0, \vcurr, 0} = \false) * (\markof{s_1, \vcurr, 0} = \true) * \procProph(\vpvs) * \cdots}$@\label{delete-proof-marknode-success}@
      $\annot{\procProph(\vpvs) * \atomicUpdate(\Phi) * \dSpec{del}(k, \abs{s_0}, \abs{s_1}, \true) * \cdots}$
      $\annot{\procProph(\vpvs) * \Phi * \historyinv(H_0[T_0+1 \rightarrowtail s_1], T_0+1) * \resources(s_1) * \cdots}$
      traverse $\vparr$ $\vsarr$ $k$; $\true$
      $\annot{\procProph(\vpvs) * \Phi}$@\label{delete-proof-post2}@
    | Failure -> 
      $\annot{\pastt{s, t_0}(\markof{s, \vcurr, 0} = \false) * (\markof{s_0, \vcurr, 0} = \true) * \cdots}$@\label{delete-proof-marknode-failure}@
      $\annot{(\exists\; s_1\; s_2\; t,\; (t_0 \leq t) * \paststate(s_1, t) * \paststate(s_2, t+1) * (\markof{s_1, \vcurr, 0} = \false) * (\markof{s_2, \vcurr, 0} = \true)) \cdots }$@\label{delete-proof-temporal-interpolation}@
      $\annot{(t_0 \leq t) * \paststate(s_1, t) * \paststate(s_2, t+1) * (\keyof{s_2, c} \notin \abs{s_2}) \cdots }$
      $\annot{\pastt{s_2, t_0}(\dSpec{del}(k, \abs{s_2}, \abs{s_2}, \false)) \cdots }$
      $\false$
      $\annot{\neg \procProph(pvs) * \pastlin(\code{del}, k, \false, t_0)}$@\label{delete-proof-post3}@
  \end{lstlisting}

\caption{Outline for the proof of the $\code{delete}$.}
\label{fig-delete-proof}
\end{figure}


}

\end{document}